\documentclass[10pt,conference]{IEEEtran}
\usepackage{cite}
\usepackage{amsmath,amssymb,amsfonts}
\usepackage{algorithmic}
\usepackage{graphicx}
\usepackage{textcomp}
\usepackage[table]{xcolor}
\usepackage[hyphens]{url}
\usepackage{fancyhdr}
\usepackage[colorlinks,
linkcolor=red,
anchorcolor=green,
citecolor=blue]{hyperref}
\usepackage{booktabs}
\usepackage{multirow}
\usepackage{graphicx}
\usepackage{bbding}
\usepackage{pifont}
\usepackage[ruled,vlined,linesnumbered]{algorithm2e}
\usepackage{setspace} 
\usepackage{threeparttable}

\SetCommentSty{mycommfont}

\SetNlSty{}{}{:}

\newcommand{\hpcayear}{2025}

\newcommand{\rOneFC}[1]{\textcolor{black}{#1}} 
\newcommand{\rTwoFC}[1]{\textcolor{black}{#1}} 
\newcommand{\rg}[1]{\textcolor{black}{#1}} 
\newcommand{\ms}[1]{\textcolor{black}{#1}} 
\newcommand{\msr}[1]{\textcolor{black}{#1}}
\newcommand{\rms}[1]{\textcolor{black}{#1}} 
\newcommand{\as}[1]{\textcolor{black}{#1}} 
\newcommand{\mv}[1]{\textcolor{black}{#1}}
\newcommand{\mvr}[1]{\textcolor{black}{#1}}

\newcommand{\rThreeFC}[1]{\textcolor{black}{#1}} 

\newcommand{\hpcasubmissionnumber}{397}
\title{Anda: Unlocking Efficient LLM Inference \mv{with a Variable-Length Grouped} 
Activation Data Format}

\def\hpcacameraready{} 
\newcommand\hpcaauthors{Chao Fang$^{\dagger,\ddagger}$, Man Shi$^{\ddagger,*}$, Robin Geens$^\ddagger$, Arne Symons$^\ddagger$, Zhongfeng~Wang$^{\dagger,*}$, Marian Verhelst$^\ddagger$}
\newcommand\hpcaaffiliation{$^\dagger$School of Electronic Science and Engineering, Nanjing University, China\hspace{1em} $^\ddagger$ESAT-MICAS, KU Leuven, Belgium}
\newcommand\hpcaemail{Email: fantasysee@smail.nju.edu.cn, zfwang@nju.edu.cn, \{man.shi, robin.geens, arne.symons, marian.verhelst\}@kuleuven.be}

\author{
  \ifdefined\hpcacameraready
    \IEEEauthorblockN{\hpcaauthors{}}
      \IEEEauthorblockA{
        \hpcaaffiliation{} \\
        \hpcaemail{}
      }
  \else
    \IEEEauthorblockN{\normalsize{HPCA \hpcayear{} Submission
      \textbf{\#\hpcasubmissionnumber{}}} \\
      \IEEEauthorblockA{
        Confidential Draft \\
        Do NOT Distribute!!
      }
    }
  \fi 
}

\fancypagestyle{camerareadyfirstpage}{%
  \fancyhead{}
  
  \fancyhead[C]{
    \ifdefined\aeopen
    \parbox[][12mm][t]{13.5cm}{\hpcayear{} IEEE International Symposium on High-Performance Computer Architecture (HPCA)}    
    \else
      \ifdefined\aereviewed
      \parbox[][12mm][t]{13.5cm}{\hpcayear{} IEEE International Symposium on High-Performance Computer Architecture (HPCA)}
      \else
      \ifdefined\aereproduced
      \parbox[][12mm][t]{13.5cm}{\hpcayear{} IEEE International Symposium on High-Performance Computer Architecture (HPCA)}
      \else
      \parbox[][0mm][t]{13.5cm}{\hpcayear{} IEEE International Symposium on High-Performance Computer Architecture (HPCA)}
    \fi 
    \fi 
    \fi 
    \ifdefined\aeopen 
      \includegraphics[width=12mm,height=12mm]{ae-badges/open-research-objects.pdf}
    \fi 
    \ifdefined\aereviewed
      \includegraphics[width=12mm,height=12mm]{ae-badges/research-objects-reviewed.pdf}
    \fi 
    \ifdefined\aereproduced
      \includegraphics[width=12mm,height=12mm]{ae-badges/results-reproduced.pdf}
    \fi
  }
  \fancyfoot[C]{}
}
\begin{document}
\maketitle

\ifdefined\hpcacameraready 
  \thispagestyle{camerareadyfirstpage}
  \pagestyle{empty}
\else
  \thispagestyle{plain}
  \pagestyle{plain}
\fi

\newcommand{\hpcaheight}{0mm}
\ifdefined\eaopen
\renewcommand{\hpcaheight}{12mm}
\fi

\renewcommand{\thefootnote}{}
\footnotetext{$^*$Corresponding author}

\begin{abstract}

The widely-used, weight-only quantized large language models (LLMs), which leverage low-bit integer \rThreeFC{(INT)} weights and retain floating-point (FP) activations, reduce storage requirements while maintaining accuracy. However, this shifts the energy and latency bottlenecks 
towards the FP activations that are associated with costly memory accesses and computations. 
\rThreeFC{Existing LLM accelerators focus \rms{primarily} 
on computation optimizations, overlooking the potential of jointly optimizing FP computations and data movement, particularly for the dominant FP-INT GeMM operations in LLM inference.}

\as{
\rThreeFC{To address these challenges,}
we investigate the sensitivity of activation precision across various LLM modules and its impact on overall model accuracy. 
Based on 
\rg{our findings}, we first propose the Anda data type: an adaptive data format with 
group-shared exponent bits and dynamic mantissa bit allocation. Secondly, we develop an iterative post-training adaptive precision search algorithm that optimizes the bit-width for different LLM modules to balance model accuracy, energy efficiency, and inference speed. Lastly, a suite of hardware optimization techniques is proposed to
\rg{maximally exploit}
the benefits of the Anda format. 
These include a bit-plane-based data organization scheme, Anda-enhanced processing units with bit-serial computation, and a runtime bit-plane Anda compressor to \ms{simultaneously} optimize storage, computation, and memory footprints.}
\rThreeFC{Our evaluations on FP-INT GeMM operations show that Anda achieves a 2.4$\times$ speedup, 4.0$\times$ area efficiency, and 3.1$\times$ energy efficiency improvement on average for popular LLMs including OPT, LLaMA, and LLaMA-2 series over the GPU-like FP-FP baseline.}
Anda demonstrates strong adaptability across various application scenarios, accuracy requirements, and system performance, enabling efficient LLM inference across a wide range of deployment scenarios.

\end{abstract}

\section{Introduction} \label{sec:intro}
Large language models (LLMs) \cite{chowdhery2023palm, zhang2022opt, touvron2023llama, team2024gemma, achiam2023gpt} have demonstrated remarkable proficiency in a wide array of natural language processing tasks, including text generation, question answering, and automatic summarization.
The extraordinary success of LLMs can be attributed to the scaling law \cite{bahri2024explaining}, which posits that performance improves dramatically with 
the ever-increased model size, training data volume, and computational resources. 
The evolution of the GPT series \cite{achiam2023gpt, kalyan2023survey, bhattacharya2024demystifying} on model size strikingly illustrates the scaling law: while GPT-1 comprised a modest 117 million parameters, its successor GPT-4 is speculated to encompass over a trillion parameters.
However, the exponential growth in LLM model sizes has created substantial deployment challenges, imposing enormous demands on storage and computational resources. 

\rOneFC{To address these challenges, quantization techniques \cite{dai2021vs, lin2024qserve, dettmers2022gpt3, lin2024awq, shao2024omniquant, frantar2023optq, xia2024fp6} have been widely adopted in LLMs to reduce memory footprint and lower deployment costs.}
\rOneFC{To maximally shrink the model size, the most common strategy for LLMs today is weight-only quantization~\cite{park2024lutgemm, dettmers2023case, chee2024quip, lin2024awq, shao2024omniquant, frantar2023optq, li2024norm, jeon2022mr, jeon2020biqgemm, xia2024fp6, xu2024onebit}, which aggressively lowers the precision of the weights while maintaining high precision for activations due to the presence of outliers \cite{dettmers2022gpt3, wei2022outlier}.}
\rOneFC{In particular, the widely adopted W4A16 scheme~\cite{shao2024omniquant, frantar2023optq}, which quantizes weights into 4-bit integers (INT4) while retaining activations in 16-bit floating-point format (FP16), significantly reduces memory requirements and bandwidth, lowering GPU memory usage by nearly 4$\times$~\cite{yuan2024llm} and facilitating deployment to smaller devices~\cite{lin2024awq, nvidia2024cutlass}.}

\begin{figure}[tbp]
    \centering
    \includegraphics[width=\linewidth]{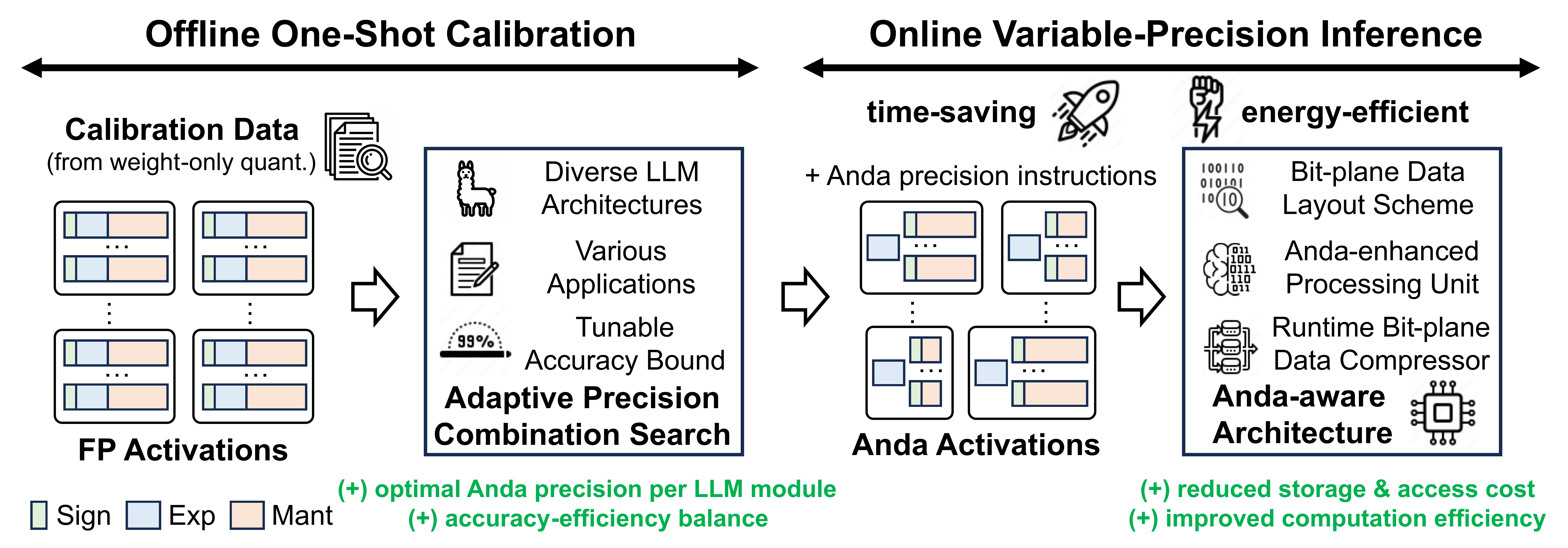}
    \caption{\rThreeFC{Overview of the drop-in replacement for FP activations using the variable-length grouped Anda data type via a \msr{one-shot} offline 
    calibration process. 
    This enables online variable-precision LLM inference,
    significantly improving speed and energy efficiency through the adaptive precision combination search algorithm and the Anda-aware architecture.}}
    \label{fig:anda_concept}
    \vspace{-2ex}
\end{figure}

\begin{figure}[tbp]
    \centering
    \includegraphics[width=\linewidth]{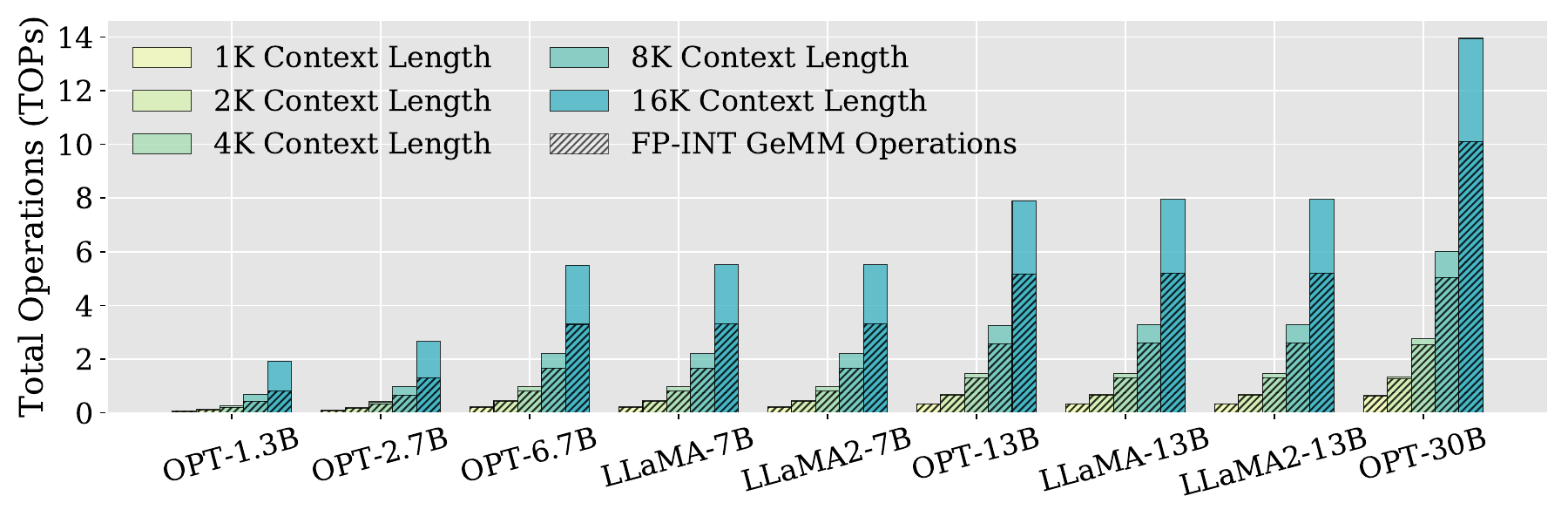}
    \caption{\rThreeFC{Proportion of FP-INT GeMM operations in weight-only quantized LLMs across varying model sizes and context lengths for text generation tasks.
    FP-INT GeMMs dominate (\textgreater 90\%) in prevalent sub-4K token applications and remain significant for 10K+ sequences.}}
    \label{fig:fp_int}
    \vspace{-3ex}
\end{figure}

\rOneFC{With the increasing importance of weight-only quantization, FP-INT GeMM operations~\cite{jang2024figna} have become indispensable in LLM inference.}
\rThreeFC{As illustrated in Fig.~\ref{fig:fp_int}, FP-INT GeMMs constitute a significant portion of the computational workload across various weight-only quantized LLMs and context lengths in text generation tasks. They dominate in typical applications with sequences under 4K tokens, comprising over 90\% of operations on average, and remain substantial even for context lengths exceeding 10K tokens in applications from LongBench \cite{bai2024longbench}.}
\rOneFC{Such prevalence highlights the urgent need to optimize FP-INT GeMMs for efficient LLM inference.}

NVIDIA’s new FP-INT GeMM kernel~\cite{nvidia2024cutlass}, as well as some specialized GPU kernels~\cite{jeon2020biqgemm, xia2024fp6, frantar2023optq, ma2024efficient}, are a consequence of this evolution.
However, these optimized GPU kernels still rely on using FP units by converting the INT weights to FP values for execution in FP GeMM operators~\cite{lin2024qserve}. 
\rOneFC{Efforts have been made to develop dedicated FP-INT arithmetic units~\cite{jang2024figna}, while the additional costs of exponent alignment and normalization persist, resulting in complicated hardware implementation.}
\rOneFC{To reduce hardware cost,}
another optimization approach is to convert the FP activations to a block floating point (BFP) data format for computation~\cite{koster2017flexpoint, drumond2018training, darvish2020pushing}.
Since grouped BFP elements share an exponent, the overhead of exponent alignment and normalization within a group disappears, simplifying the operation to \rOneFC{INT} 
arithmetic. 
However, to mitigate accuracy loss when converting FP activations to BFP format on a pre-trained network, costly retraining \cite{dai2021vs, darvish2023shared, darvish2020pushing, keller202395, zhang2022fast, koster2017flexpoint, noh2023flexblock, guo2023boost} is required, hindering agile LLM deployment. Alternatively, accuracy can be preserved by using a large mantissa field conversion~\cite{jang2024figna, kim2023winning}, but this significantly increases the energy consumption due to computational and memory access overhead of the additional bits.


\rOneFC{In summary, processing
FP activations remains a major bottleneck in weight-only quantized LLM inference, and existing methods struggle to balance model accuracy, computational efficiency, and energy consumption.}
\as{
\rTwoFC{To overcome the above limitations, we propose Anda to unlock efficient LLM inference. Anda introduces a novel variable-length grouped activation data format, coupled with 
\rTwoFC{the algorithm innovation}
and specialized hardware optimizations.}
\rThreeFC{As shown in Fig.~\ref{fig:anda_concept}, Anda first employs a fast, training-free adaptive precision search algorithm \msr{during compile time,}
using the same calibration data as post-training weight-only quantization~\cite{gong2024llm}. 
\msr{Guided by user-defined accuracy constraints, our one-shot process identifies the desired mantissa bit length, which instructs the activation precision across various LLM modules during inference.}
Combining the \msr{flexible}
Anda \msr{data} format with our specialized hardware architecture \msr{allows running these dominant FP-INT GeMM operations at lower precision,}
significantly improving inference speed and energy efficiency for weight-only quantized LLMs.}}
More concretely, our contributions are as follows:
\begin{itemize}
    \item We investigate the potential of BFP activation quantization across popular LLM models within different modules. Based on these insights, we propose Anda: a variable-length grouped data format with shared exponents and adjustable mantissa widths for activations.
    \item We develop an adaptive search algorithm to optimize the mantissa widths for different LLM modules without retraining. 
    \rTwoFC{This algorithm balances model accuracy, energy efficiency, and inference speed based on a user-defined accuracy loss tolerance.}
    \item We design 
    an efficient Anda-aware hardware architecture
    featuring (a) a bit-plane-based data layout scheme in~memory, (b) Anda-enhanced bit-serial processing units, and (c) a runtime bit-plane data compressor. 
    \rThreeFC{Extensive evaluations of the Anda system across popular LLMs demonstrate an average improvement in processing speed of 2.4$\times$, area efficiency of 4.0$\times$ and energy efficiency of 3.1$\times$ compared to existing SotA hardware.}
\end{itemize}


The paper is organized as follows: Sec.~\ref{sec:bkg} reviews the~benefits and remaining bottlenecks of weight-only quantized~LLMs and BFP formats, and quantifies the sensitivity of LLM~inference accuracy to the shared exponents and reduced mantissas sizes. Based on these findings, Sec.~\ref{sec:algo} features the proposed Anda data format and presents the algorithm to rapidly optimize mantissa length under given accuracy constraints. Next, an Anda-optimized hardware architecture is proposed in Sec.~\ref{sec:arch}, which allows us to derive system-level gains in Sec.~\ref{sec:eval} and benchmark it against the SotA solutions. Sec.~\ref{sec:concls} concludes the paper.

\section{Background and Motivation}\label{sec:bkg}

\subsection{Weight-only Quantized LLMs} \label{subsec:weight-only}

\rOneFC{Weight-only quantization~\cite{park2024lutgemm, dettmers2023case, chee2024quip, lin2024awq, shao2024omniquant, frantar2023optq, li2024norm, jeon2022mr, jeon2020biqgemm, xia2024fp6, xu2024onebit} has emerged as a pivotal technique for efficient LLM inference.}
\rOneFC{Unlike weight-activation quantization~\cite{dai2021vs, lin2024qserve, dettmers2022gpt3, xiao2023smoothquant, zhao2024atom}, which reduces precision for both weights and activations, weight-only quantization focuses solely on compressing model parameters using a much more aggressive quantization scheme.}

\begin{figure}[tbp]
    \centering
    \includegraphics[width=0.96\linewidth]{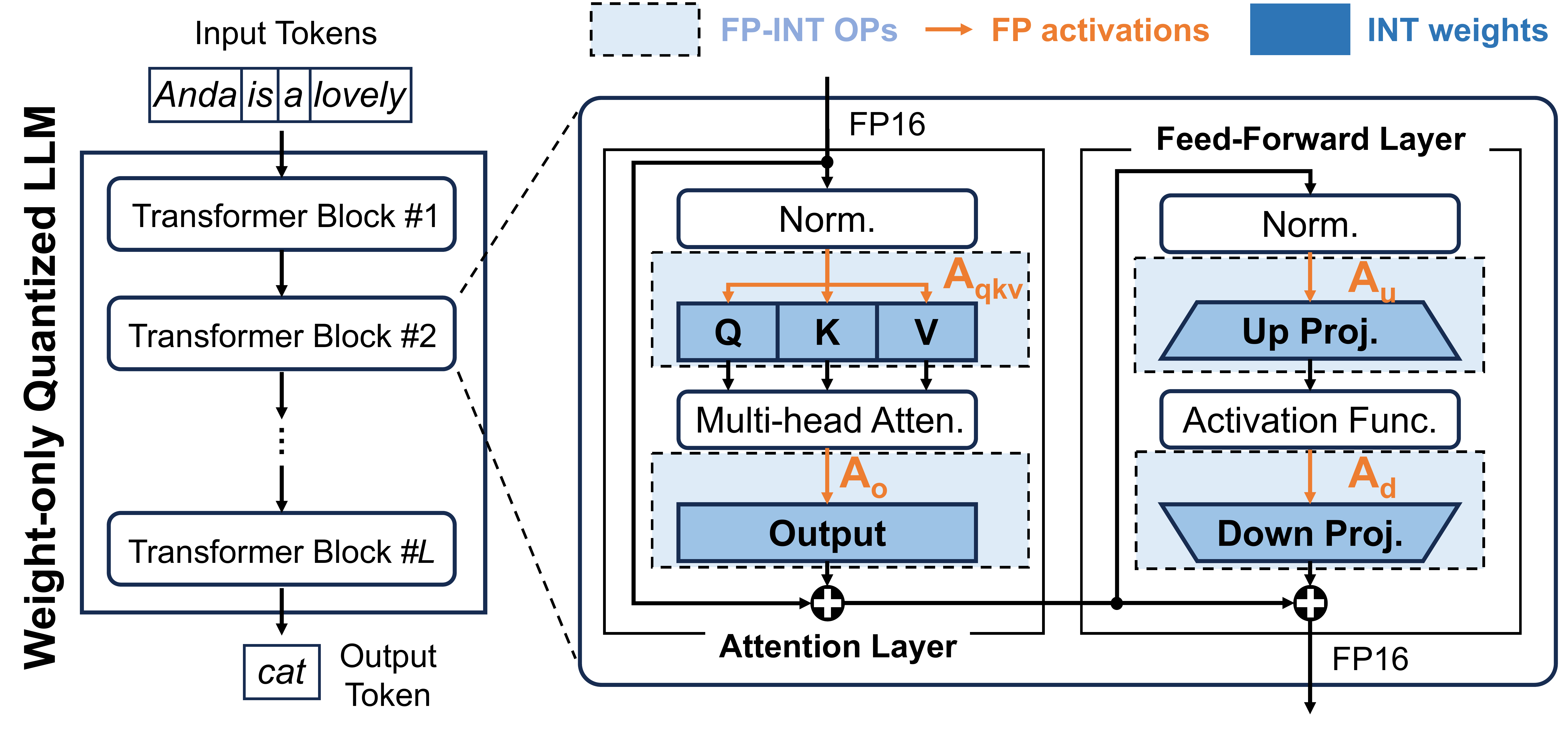}
    \caption{Illustration of the architecture for a weight-only quantized LLM model.}
    \label{fig:llm_blocks}
    \vspace{-3ex}
\end{figure}

\rOneFC{Fig.~\ref{fig:llm_blocks} illustrates the architecture of a weight-only quantized LLM, composed of a series of Transformer blocks that each contains an attention layer and a feed-forward layer.}
\rOneFC{The light blue background highlights the dominant computational modules involving FP-INT GeMM operations, which can be categorized into four module types based on the positions of the FP activations: the first type involves $A_{qkv}$ interacting with $W_q$, $W_k$, and $W_v$ to compute the query ($Q$), key ($K$), value ($V$) matrices, respectively; the second type involves $A_o$ multiplying with $W_o$ to compute the output matrix;}
\rTwoFC{the other two types are up-projection and down-projection modules of the feed-forward layer, respectively, involving $A_u$ and $A_d$ with interacting to corresponding weights.}
\rThreeFC{Weight-only quantized LLMs offer significant advantages in storage efficiency~\cite{nvidia2024cutlass,shao2024omniquant}.}
\rThreeFC{Compared to W8A8 weight-activation quantized LLMs~\cite{xiao2023smoothquant}, W4A16 weight-only quantized LLMs~\cite{lin2024awq} achieve similar model accuracy while reducing storage requirements \msr{of model parameters} by nearly half~\cite{yuan2024llm}, making them particularly suitable for deployment on resource-constrained devices in edge computing scenarios.}
\rThreeFC{However, under current GPU computing schemes, computing a W4A16 FP-INT operations consumes approximately 1.7$\times$ more energy than W8A8 INT-only operations~\cite{kim2023winning}.}
\rThreeFC{This can be explained by accessing FP activations incurs higher energy costs than INT weights~\cite{huang2024isca}, and FP-INT operations require complicated hardware implementations~\cite{jang2024figna}.}
\rOneFC{Hence, optimizing FP activations emerges as a key opportunity to improve the overall efficiency of weight-only quantized LLMs.}

\begin{figure}[tbp]
    \centering
    \includegraphics[width=0.96\linewidth]{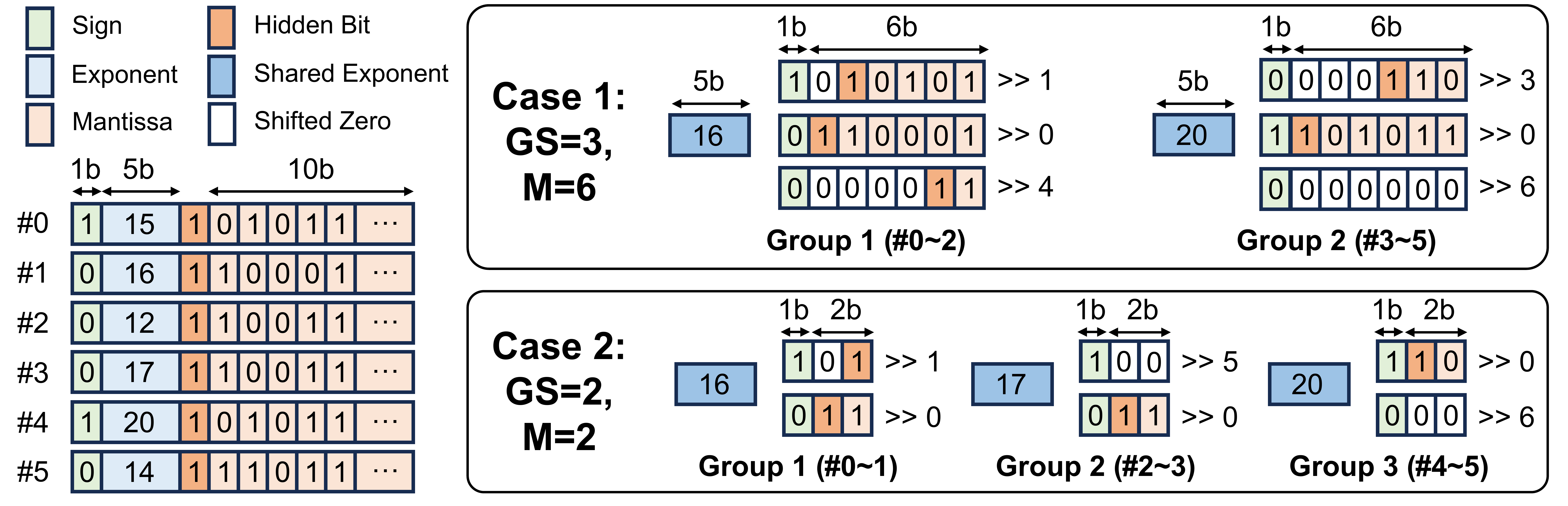}
    \caption{The process of converting a set of FP16 numbers into different BFP numbers. BFP format is regulated by two key parameters: group size (GS) and mantissa length (M).}
    \label{fig:fp_bfp_convert_process}
    \vspace{-1em}
\end{figure}

\subsection{Block Floating Point}

Reducing the computation and storage overhead of FP16 activations is crucial for optimizing the efficiency of LLMs.
\rOneFC{BFP~\cite{drumond2018training} offers a promising solution by sharing exponents within groups of values, preserving dynamic range while mitigating the impact of outliers and simplifying computations.}
\rOneFC{The BFP format can characterized by two key parameters: group size and mantissa length.}
\rOneFC{Fig.~\ref{fig:fp_bfp_convert_process} shows the process of converting FP16 tensors to BFP numbers using two different instances of the BFP format.}
\rOneFC{Initially, FP16 tensors are divided into groups. Within each group, the largest exponent is selected as the shared exponent and other mantissas are right-shifted based on their exponent differences. Bits exceeding the specified mantissa length are truncated, and zero is represented by all mantissa bits being 0.}
\rOneFC{As illustrated in Fig.~\ref{fig:fp_bfp_convert_process}, this conversion process can lead to precision loss due to mantissa truncation, with some elements becoming zero, thereby posing a significant challenge to maintaining model accuracy.}

\rOneFC{Current approaches to address this fall into two categories.}
On the one hand, BFP-aware training fine-tunes the model after the quantization~\cite{dai2021vs, darvish2023shared, darvish2020pushing, keller202395, zhang2022fast, koster2017flexpoint, noh2023flexblock, guo2023boost, kim2024isca, fan2019static}, at the expense of a
costly training process, making it rather impractical for agile LLM deployment. 
\rTwoFC{On the other hand, direct conversion of pre-trained FP models to BFP formats~\cite{fan2018reconfigurable, lian2019high, koster2017flexpoint, fan2019static, noh2023flexblock} requires long mantissas to avoid the significant accuracy loss, which increases computation and storage overhead, diminishing the advantages of BFP.}
\rOneFC{To avoid the storage of these long mantissas, methods like FIGNA~\cite{jang2024figna} and~\cite{kim2023winning} propose dynamic conversion to BFP during computation.
This approach stores activations in FP16 format and expands to long mantissas with shared exponents before computations to maintain model accuracy.}
However, this also prevents FIGNA from obtaining activation memory footprint savings.

To avoid both costly retraining and large activation memory footprints, we seek a solution that can rapidly convert FP activations to BFP activations without retraining, while also leveraging the computational and storage advantages of BFP for LLM inference.
To achieve this goal, it is necessary to explore opportunities for reduced mantissa length BFP under the unique characteristics of LLMs.

\subsection{Opportunities towards Activation Optimizations}

\rOneFC{We explore opportunities for LLM activation optimization by investigating the sensitivity of model accuracy to reduced mantissa lengths in BFP formats. This study converts FP-INT GeMM activation tensors ($A_{qkv}$, $A_{o}$, $A_{u}$, $A_{d}$) from FP16 to BFP format, as shown in Fig.~\ref{fig:fp_bfp_convert_process}. Model accuracy is evaluated using perplexity (PPL) on the WikiText2 dataset, with lower PPL indicating higher accuracy. We assume a 1\% accuracy loss tolerance in practical scenarios. 
We aim to uncover efficient activation representations while maintaining LLM performance within acceptable limits.}


\begin{figure}[tbp]
    \centering
    \includegraphics[width=\linewidth]{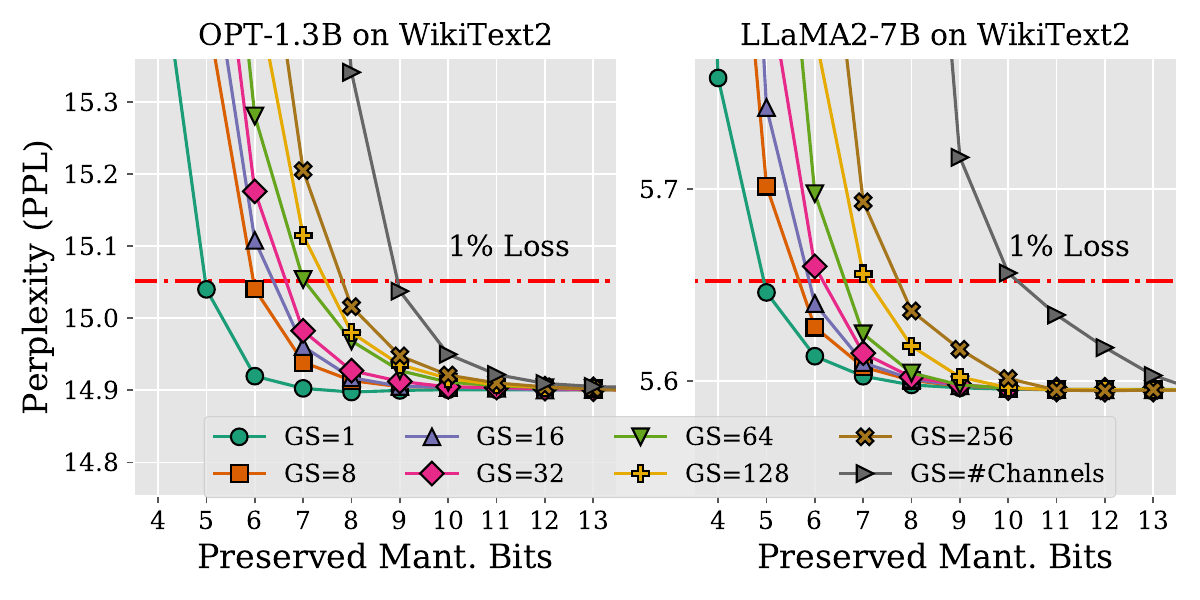}
    \caption{LLM sensitivity to BFP group size (GS) and preserved mantissa bits.}
    \label{fig:acc_wrt_gs}
    \vspace{-1em}
\end{figure}

\mv{\textbf{Sensitivity to group size:}} 
\rOneFC{Fig.~\ref{fig:acc_wrt_gs} illustrates the sensitivity to shared exponent group size for two different LLM models across various mantissa lengths. The experiments reveal a clear trade-off between group size and the minimum required mantissa length to maintain model accuracy.}
\rOneFC{Larger activation group sizes allow more efficient parallel computations, yet at a greater accuracy tolerance or increased mantissa lengths.}
\rOneFC{Based on these observations, we select a group size of 64 for subsequent experiments, as it offers a good balance between computational efficiency and accuracy tolerance.}

\begin{figure}[tbp]
    \centering
    \includegraphics[width=\linewidth]{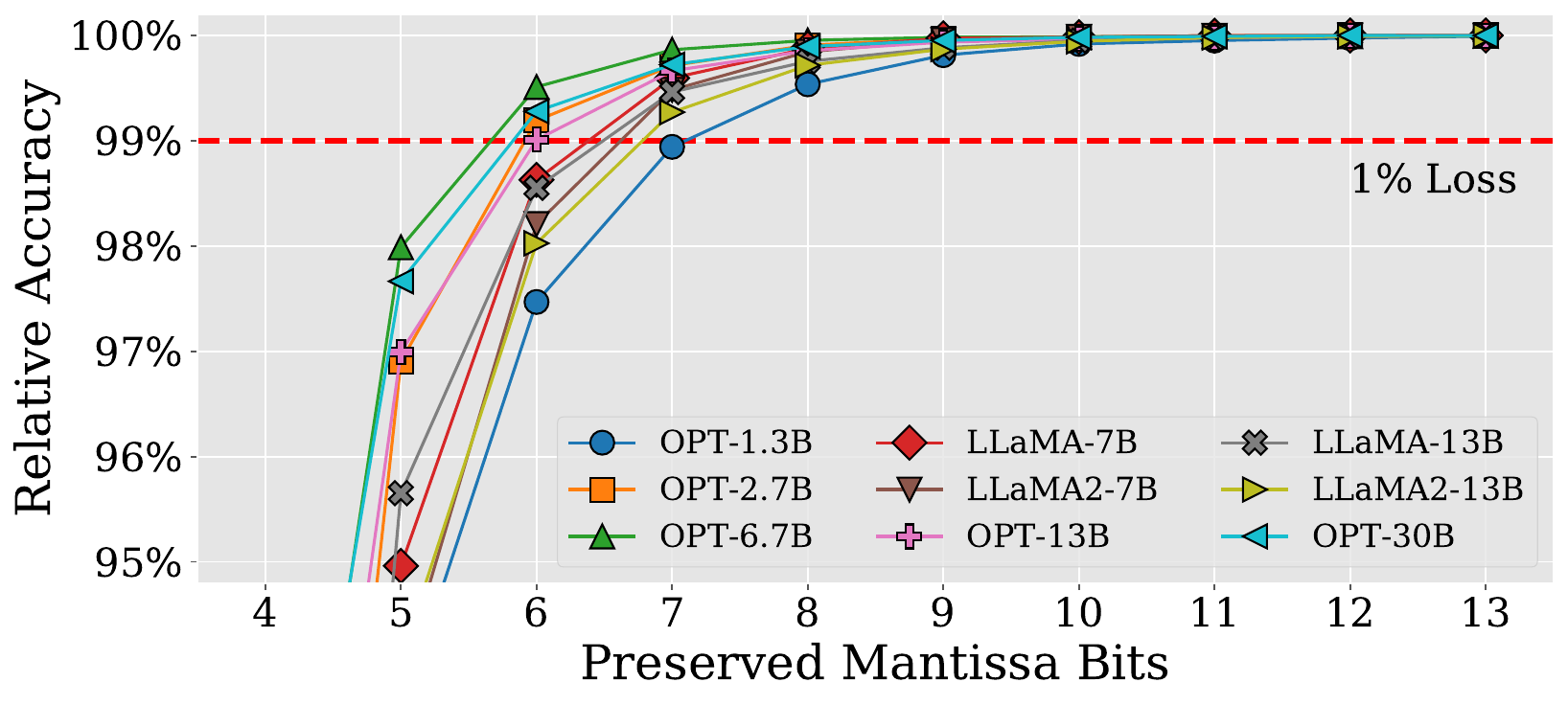}
    \caption{The relative accuracy to preserved mantissa bits across various LLMs.}
    \label{fig:acc_wrt_mbits}
    \vspace{-1em}
\end{figure}

\mv{\textbf{Sensitivity to LLM model:}} With this group size of 64, we continue our exploration across a wider range of recent LLMs, to derive their sensitivity to reduced mantissa lengths.
\rOneFC{Fig.~\ref{fig:acc_wrt_mbits} reveals varying sensitivities among different models.}
Notably, models such as OPT-2.7B, OPT-6.7B, OPT-13B, and OPT-30B are less sensitive to mantissa reduction, allowing for the direct removal of 5 mantissa bits, while other models could only tolerate the removal of 4 mantissa bits. 
\rOneFC{As more mantissa bits are removed, differences in accuracy sensitivity become more pronounced.}
This insight inspires us to consider a variable-length BFP datatype, potentially enabling more aggressive compression in less sensitive models while employing a more conservative one for others. 
\rOneFC{It also prompts us to explore whether activations in different modules within one LLM have varying sensitivities.}

\begin{figure}[tb]
    \centering
    \includegraphics[width=\linewidth]{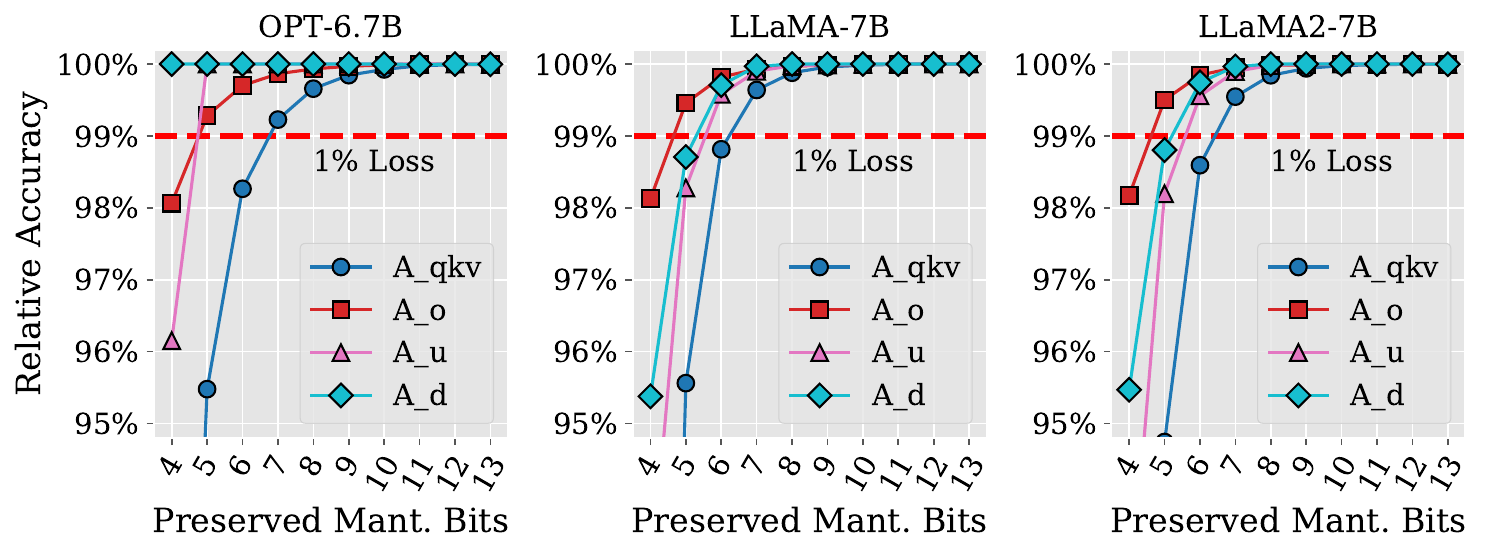}
    \caption{The relative accuracy of OPT-6.7B, LLaMA-7B, and LLaMA2-7B when cutting mantissa bits on either $A_{qkv}$, $A_{o}$, $A_{u}$, or $A_{d}$ activation only.}
    \label{fig:module_wise_acc}
    \vspace{-1em}
\end{figure}

\mv{\textbf{Sensitivity to LLM inner module:}}
\rOneFC{We finally explore the impact of different mantissa lengths of the activations of different modules within the same LLM. More specifically, we examine the $A_{qkv}$, $A_{o}$, $A_{u}$, and $A_{d}$ modules of the OPT-6.7B, LLaMA-7B, and LLaMA2-7B models. The mantissa length of each module is swept while keeping the lengths of other modules fixed at 13 bits. Fig.~\ref{fig:module_wise_acc} summarizes the results, revealing that activations from different modules have varying impacts on model accuracy across all three models. $A_{qkv}$ consistently shows the most significant influence, while $A_{d}$ demonstrates low sensitivity in OPT-6.7B but has a more pronounced effect in the LLaMA series models.}

\rOneFC{Our study reveals several key insights into the application of BFP in LLMs:
(a) LLMs can maintain good performance with reduced mantissa lengths. 
(b) Different LLM models exhibit varying sensitivities to mantissa reduction. 
(c) Within a single LLM, different modules have distinct sensitivities to precision reduction.
These observations motivate us to introduce the new variable-length grouped data format for FP activations, along with a methodology for post-training quantization (PTQ) and rapid selection of tolerable reduced mantissa lengths for any LLM.}



\section{Anda Data Format} \label{sec:algo}
In this section, we present unique features of the Anda data format and demonstrate 
\rg{its benefits towards FP-INT operations}
in weight-only quantized LLM inference. Furthermore, we introduce 
a \ms{mantissa bit-width search} method
to efficiently identify the optimized Anda precision combinations that satisfy a user-defined accuracy drop.

\subsection{Anda Format Features}


\rTwoFC{
\rg{Based on the} findings of our previous study, we propose the Anda format: an innovative variable-length mantissa BFP scheme designed for efficient LLM inference.}
\rTwoFC{Anda's structure comprises a sign bit, a shared exponent, and a variable-length mantissa, building upon traditional BFP conversion processes as previously shown in Fig.~\ref{fig:fp_bfp_convert_process}.}
\rTwoFC{Its key feature is the ability to dynamically select mantissa lengths for different tensors based on their precision sensitivity, maintaining consistency within each tensor while optimizing the accuracy-efficiency trade-off.}

\rTwoFC{Table~\ref{tab:bfp_cmp} compares Anda with prior BFP formats, categorizing them based on supported mantissa lengths.}
\rTwoFC{Uni-length formats, such as VS-Quant~\cite{dai2021vs} and FIGNA~\cite{jang2024figna}, use fixed mantissa lengths, while multi-length formats like FAST~\cite{zhang2022fast} and DaCapo~\cite{kim2024isca} offer limited flexibility with 2$\sim$3 predefined lengths. Anda surpasses both by providing a continuous range of mantissa lengths, allowing fine-grained precision control across different LLM modules.}
\rg{Enabled by specialized hardware units, as detailed in Sec.~\ref{sec:arch}, smaller mantissa widths result in a lower inference latency, computational cost and memory storage cost. This allows Anda format to carefully balance model precision and computational efficiency, providing a more aggressive compression in less sensitive model
parts while preserving critical precision elsewhere.
} 

\begin{table}[t]
\centering
\caption{Anda format definition in contrast with prior BFP formats}
\label{tab:bfp_cmp}
\resizebox{0.46\textwidth}{!}{%
\begin{tabular}{@{}ccccc@{}}
\toprule
                     & BFP Type                      & \begin{tabular}[c]{@{}c@{}}Mantissa Length \\ during Computation\end{tabular} & Computation                    & Storage                      \\ \midrule
VS-Quant~\cite{dai2021vs}             & \multirow{6}{*}{Uni-Length}   & 4b                                                                            & \multirow{6}{*}{Bit-parallel} & BFP Element-based            \\
BOOST~\cite{guo2023boost}                &                               & 5b                                                                            &                               & BFP Element-based            \\
X. Lian et al.~\cite{lian2019high}       &                               & 8b                                                                            &                               & BFP Element-based            \\
FIGNA~\cite{jang2024figna}                &                               & 14b                                                                           &                               & FP16 Element-based           \\
H. Fan et al.~\cite{fan2018reconfigurable}        &                               & 15b                                                                           &                               & BFP Element-based            \\
Flexpoint~\cite{koster2017flexpoint}            &                               & 16b                                                                           &                               & BFP Element-based            \\ \midrule
FAST~\cite{zhang2022fast}                 & \multirow{3}{*}{Multi-Length} & 2b/4b                                                                         & Chunk-serial                  & BFP Chunk-based              \\
DaCapo~\cite{kim2024isca}               &                               & 2b/4b/8b                                                                      & \multirow{2}{*}{Bit-parallel} & BFP Element-based            \\
FlexBlock~\cite{noh2023flexblock}            &                               & 4b/8b/16b                                                                     &                               & BFP Element-based            \\ \midrule
\rowcolor[HTML]{FFFFC7} \textbf{Anda (Ours)} & \textbf{Variable-Length}      & \textbf{1b/2b/.../16b}                                                     & \textbf{Bit-serial}           & \textbf{BFP Bit-plane-based} \\ \bottomrule
\end{tabular}%
}
\vspace{-1em}
\end{table}


\begin{figure*}[t]
    \centering
    \includegraphics[width=0.92\textwidth]{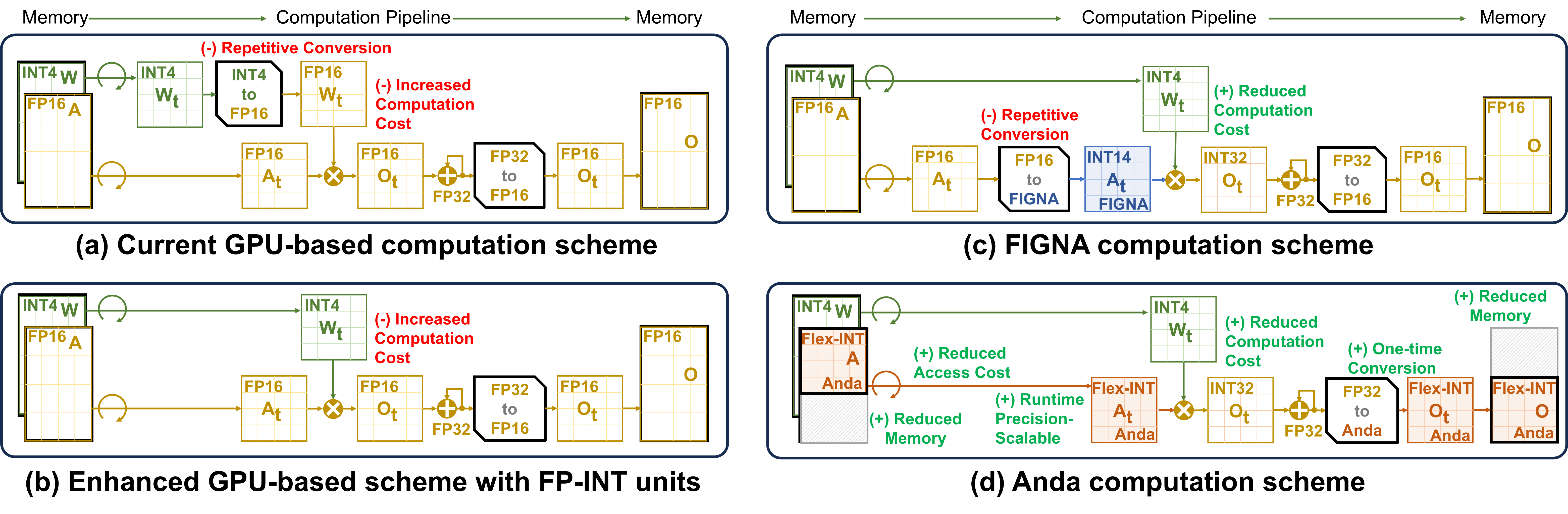}
    \caption{Comparison of (a) the current computation scheme on GPU, (b) and that enhanced with dedicated FP-INT processing unit, (c) FIGNA scheme, and (d) our Anda scheme for FP-INT GeMM. Our Anda scheme significantly reduces memory space, access cost, and computation cost and enables energy-efficient precision-scalable operations.}
    \label{fig:workflow}
    \vspace{-1em}
\end{figure*}
\subsection{Efficient FP-INT GeMM Using Anda Format}

\rOneFC{We then compare the workflows of \ms{GeMM workloads} of several SotA approaches 
to illustrate the advantages of replacing FP16 activations with the Anda data format. Taking the W4A16 quantization scheme as an example, we examine the FP-INT GeMM computation process (a) on existing GPU platforms~\cite{lin2024qserve}; (b) on GPU platforms with dedicated FP-INT processing units; (c) using FIGNA's dynamic conversion scheme~\cite{jang2024figna}; and (d) with our proposed Anda approach. Fig.~\ref{fig:workflow} depicts 
the \ms{four} schemes, with colors indicating the data types used throughout the computational process.}


\ms{Fig.~\ref{fig:workflow}(a) shows the \rg{workflow of} W4A16 LLMs on common GPU platforms}. 
\rTwoFC{The absence of dedicated FP-INT computation units in GPU necessitates converting INT4 weights to FP16 before processing, with tensor cores operating in FP16 mode.}
This scheme not only brings additional format conversion overheads, but requires costly FP computations.

\rOneFC{GPU platforms equipped with dedicated FP-INT processing units, as illustrated in Fig.~\ref{fig:workflow}(b), can eliminate the need for converting INT4 weights to FP16, thereby reducing data conversion overheads and computation costs.}
\rOneFC{However, as pointed out by FIGNA~\cite{jang2024figna}, the high alignment and normalization overhead associated with FP-INT processing units still results in high computational expenses.}

\rOneFC{To efficiently deploy W4A16 LLMs, FIGNA proposes a computation scheme using a BFP variant with corresponding hardware support to overcome the issues with dedicated FP-INT units.}
As \ms{depicted} in Fig.~\ref{fig:workflow}(c), activations are stored in FP16 format in memory, converted to the FIGNA format before computation, \mv{after which} a 14-bit \mv{mantissa is multiplied} with INT4 weights for GeMM computation. The final results are then converted \mv{again} to FP16 and written back to memory.
\rOneFC{This scheme reduces the computation overhead by converting costly FP GeMM to INT operations.}
However, since FP16 activations need to be repeatedly \mv{accessed} during computation, frequent data conversion from FP16 to FIGNA introduces additional overhead, affecting overall efficiency.


\rOneFC{As presented in Fig.~\ref{fig:workflow}(d), our proposed Anda format computation scheme offers some unique advantages in contrast with the previous approaches.}
\rOneFC{Firstly, the activations are no longer stored in memory in FP16 \rg{format}, but directly in the Anda data format, reducing storage overhead and data access overhead while avoiding frequent data conversion.}
\rOneFC{Secondly, the shared exponent enables INT dot-product operations within a group, followed by FP32 accumulation across groups, reducing the computational overhead of FP-INT GeMMs.}
\rOneFC{Thirdly, the variable-length mantissa \ms{considerably decreases} dot-product operations and memory accesses use the minimal necessary word length.}
\rOneFC{Finally, converting only the final FP32 results back to Anda format before writing to memory minimizes the storage \ms{requirement} 
and the additional overhead from switching data format.}

\subsection{Adaptive Precision Combination Search}


\rThreeFC{To leverage the Anda format for fast deployment and hardware performance gains, we propose an adaptive precision search algorithm for offline compile\mvr{-time} optimization of activation \mvr{precision}s in weight-only quantized LLMs.
Our algorithm is built around two key strategies. 
(a) We narrow the search space to \mvr{the precision of only} four key tensor types 
\mvr{,i.e., $A_{qkv}$, $A_o$, $A_u$, and $A_d$},
based on their sensitivity to model accuracy as demonstrated in Fig.~\ref{fig:module_wise_acc}. \mvr{This precision} combination is represented as a 4-tuple $[M_{qkv}, M_o, M_u, M_d]$.
(b) We employ a training-free, one-shot calibration process reusing the small amount of calibration data from the post-training weight-only quantization process, being several thousands of tokens with hundred batches~\cite{frantar2023optq, lin2024awq, shao2024omniquant}.
Though prior layer-wise methods~\cite{han2017ese, wang2019haq, dong2019hawq} may achieve finer precision adjustments, their prolonged search times significantly extend the deployment process. 
In contrast, our module-wise approach rapidly assigns mantissa lengths while maintaining consistency across layers and can easily be integrated into standard post-training deployment workflows.}
\setlength{\textfloatsep}{1em} 
\begin{algorithm}[t]
\small
\SetAlgoLined
\caption{Adaptive Precision Combination Search}
\label{alg:anda_search}
\KwIn{LLM model $L$, calibration dataset $D$, \newline accuracy loss tolerance $\delta$, max iterations $N$}
\KwOut{Optimized precision combination $best\_comb$ denoted as a 4-tuple $[M_{qkv}, M_{o}, M_{u}, M_{d}]$}
\tcp{S1: Initialize search starting points}
$Q \gets PriorityQueue([4,4,4,4], ..., [13, 13, 13, 13])$\;
$best\_comb \gets \text{null}$, $best\_bops \gets \infty$\; 
\textit{iterations} $\gets$ 0, $visited \gets \{\}$\;
$fp\_acc \gets \textbf{EvaluateAccuracy}(L, D)$\; 
\While{iterations $< N$}{
    \tcp{S2: Check the promising combination}
    $bops\_eval \gets Q.\text{map}(\textbf{EvalBOPs})$\;
    $curr\_bops \gets min(bops\_eval)$\;
    $curr\_comb \gets Q.\text{get}(bops\_eval.\text{index}(curr\_bop))$\;
    $visited \gets visited \cup \{curr\_comb\}$\;
    $anda\_acc \gets \textbf{EvaluateAccuracy}(L, D, curr\_comb)$\;
    \tcp{S3: Update and relax the best combination}
    \If{$curr\_bops < best\_bops$ \textnormal{\textbf{and}} $anda\_acc \geq (1-\delta) \cdot fp\_acc$}{
        $best\_comb \gets curr\_comb$\; 
        $best\_bops \gets curr\_bops$\;
        $neighbors \gets \textbf{GenerateCandidates}(curr\_comb)$\;
        \ForEach{$n \in neighbors$}{
            \If{$n \notin visited$}{
                $Q.\text{push}(n)$\;
            }
        }
    }
    \If{$Q.\text{empty}()$}{
        \textbf{break}\;
    }
    \textit{iterations} $\gets$ \textit{iterations} + 1\;
}
\Return $best\_comb$
\end{algorithm}


\rOneFC{As outlined in Algorithm~\ref{alg:anda_search}, we take the LLM model~$L$, a calibration dataset $D$, an accuracy loss tolerance $\delta$, and a maximum number of iterations $N$ as inputs. 
The accuracy tolerance $\delta$ specifies the acceptable level of performance degradation, while the maximum number of iterations $N$ serves as a termination criterion, ensuring the algorithm concludes within a reasonable time frame. With these inputs, our algorithm finds the optimal 4-tuple precision combination within the given iterations that best balances model accuracy and inference efficiency across the model's key activation components.}
The search process consists of three key steps.

\rOneFC{\textbf{Step 1: Initialize search starting points.} 
A priority~queue with precision combinations of equal precision across all modules \ms{is initialized first}. These \ms{precision} combinations range from aggressive (e.g., $[4,4,4,4]$) to conservative (e.g., $[13,13,13,13]$). This strategy enables the rapid discovery of efficient combinations while ensuring the existence of feasible solutions, as validated by our prior experiments in Fig.~\ref{fig:acc_wrt_mbits}.}


\rOneFC{\textbf{Step 2: Check the promising combination.} 
In each iteration, the combination with the lowest bit operations (BOPs) is extracted from the priority queue and added to the visited set.
The BOP metric~\cite{abati2023resq, koryakovskiy2023one, li2024quasar, tian2023bebert} quickly estimates computational cost by calculating the total number of bit operations for the necessary multiplications under a given combination. 
This allows us to efficiently prioritize promising combinations 
without a full model evaluation. 
The accuracy of the promising combination is then examined on the calibration dataset.} 

\rOneFC{\textbf{Step 3: Update and relax the best combination.} If the evaluated combination yields lower BOPs than the current best while maintaining accuracy within the specified tolerance, it becomes the new best combination. 
To generate nearby precision candidates, the algorithm then relaxes this best combination by decreasing the mantissa length of each tensor type by one, while keeping the other tensor types unchanged.
For example, if the current best combination is $[6,7,5,5]$, the generated candidates will be $[5,7,5,5]$, $[6,6,5,5]$, $[6,7,4,5]$, and $[6,7,5,4]$.
The generated candidates that have not been visited before are added to the priority queue.
If the accuracy constraint is not met, no update is made.}
Step 2 and 3 are repeated until the maximum number of iterations is reached or the search space is exhausted.



\begin{figure}[tbp]
    \centering
    \includegraphics[width=0.95\linewidth]{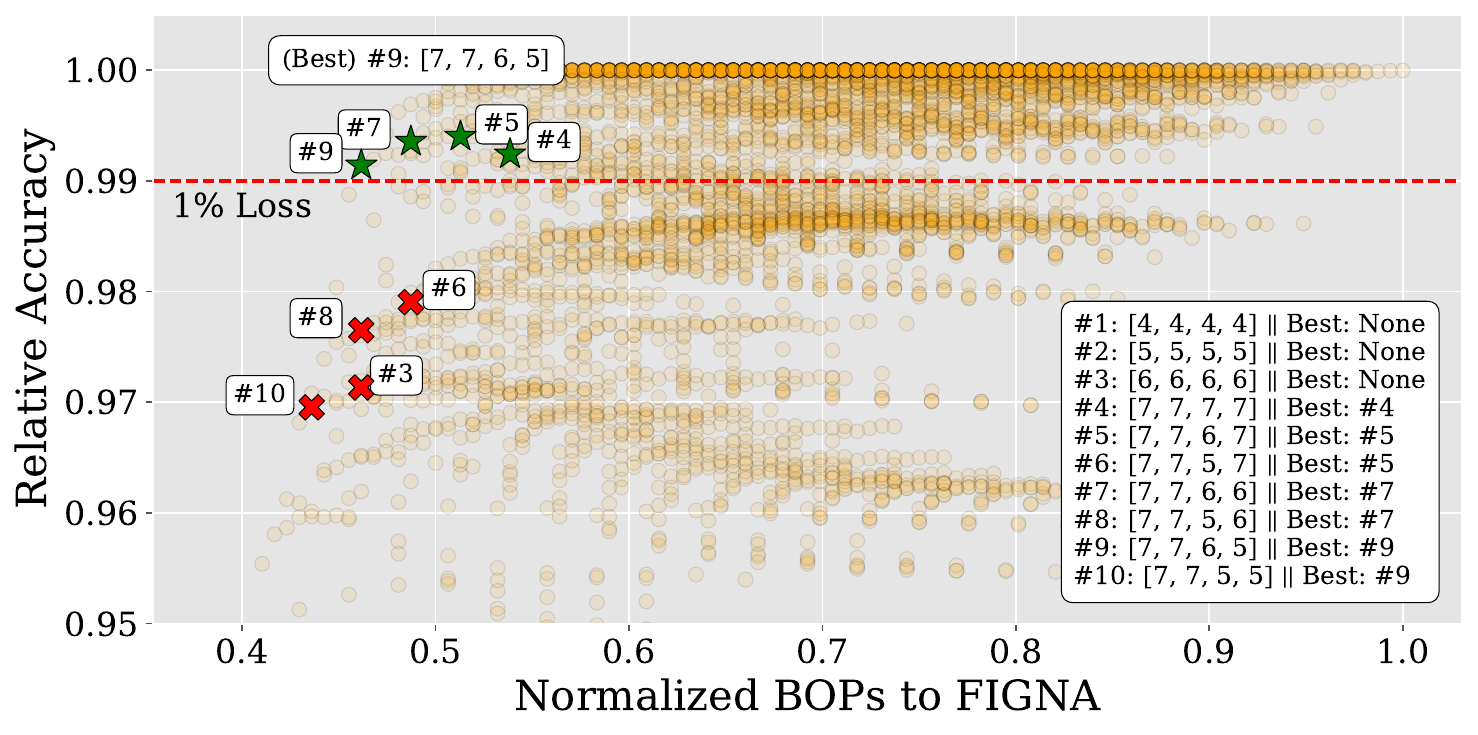}
    \caption{Search process of the proposed adaptive precision combination search algorithm on the OPT-125M model with constraint under 1\% accuracy loss, which efficiently finds the global optimum within 10 iterations.}
    \label{fig:prec_comb_eff}
\end{figure}

\subsection{Precision Combination Search Efficiency} \label{subsec:pre_comb}

\rThreeFC{Our algorithm aims to efficiently optimize FP activations in weight-only quantized LLMs during the post-training phase. Most weight-only quantization processes~\cite{lin2024awq, frantar2023optq, shao2024omniquant} rely on a small calibration dataset, which we can reuse \mvr{in the activation precision search.} 
Ensuring a rapid search process is critical to avoid extending post-training deployment time. 
Therefore, the algorithm is designed to find a near-optimal solution quickly, within an acceptable accuracy tolerance, to enable efficient hardware deployment.}


\rThreeFC{The efficiency of our algorithm is enhanced by two key mechanisms: First, we introduce a constraint that updates the best combination only when a new \mvr{precision combination} offers a lower computational cost, employing a relaxation strategy similar to gradient descent to accelerate convergence. While this may miss the global optimum, it ensures a high-performance combination within limited iterations. Second, we set an iteration limit to complete the search within a reasonable timeframe, avoiding deployment delays. \mvr{It is here important to note that the relatively limited search space of only 4 precision variables allows for fast convergence with just a few iterations.}
The execution time of each iteration is roughly the time of a forward pass over the calibration dataset to validate the precision combination.}

\rThreeFC{To demonstrate our algorithm's search efficiency, we compare it with the conventional brute-force approaches~\cite{dai2021vs,darvish2020pushing,darvish2023shared} on the OPT-125M model.}
\rThreeFC{As shown in Fig.~\ref{fig:prec_comb_eff}, the search space for OPT-125M contains over 10,000 possible combinations, and our algorithm identifies the precision combination $[7,7,6,5]$ in just 10 iterations, maintaining accuracy within 1\% loss.}
\rThreeFC{In practice, we limit the search to 32 iterations, ensuring that time overhead remains minimal while achieving a near-optimal precision combination. By avoiding time-consuming backward propagation or complex solving processes, our algorithm operates approximately twice as fast as Omniquant~\cite{shao2024omniquant} and ten times faster than GPTQ~\cite{frantar2023optq}, the current SoTA methods for post-training weight-only LLM quantization.}

\section{Anda Architecture} \label{sec:arch}

\rTwoFC{In this section, we first present the three key components of the Anda architecture: (a) A variable-length activation data layout in on-chip memory storage, (b) an Anda-enhanced bit-serial processing unit, and (c) a runtime bit-plane compressor for output activations. 
These components collectively enhance storage efficiency, computational performance, and energy conservation. 
Finally, we present how these components integrate to form the overall Anda architecture, a computing system optimizing LLM inference using the Anda format.}


\begin{figure}[tbp]
    \centering
    \includegraphics[width=0.92\linewidth]{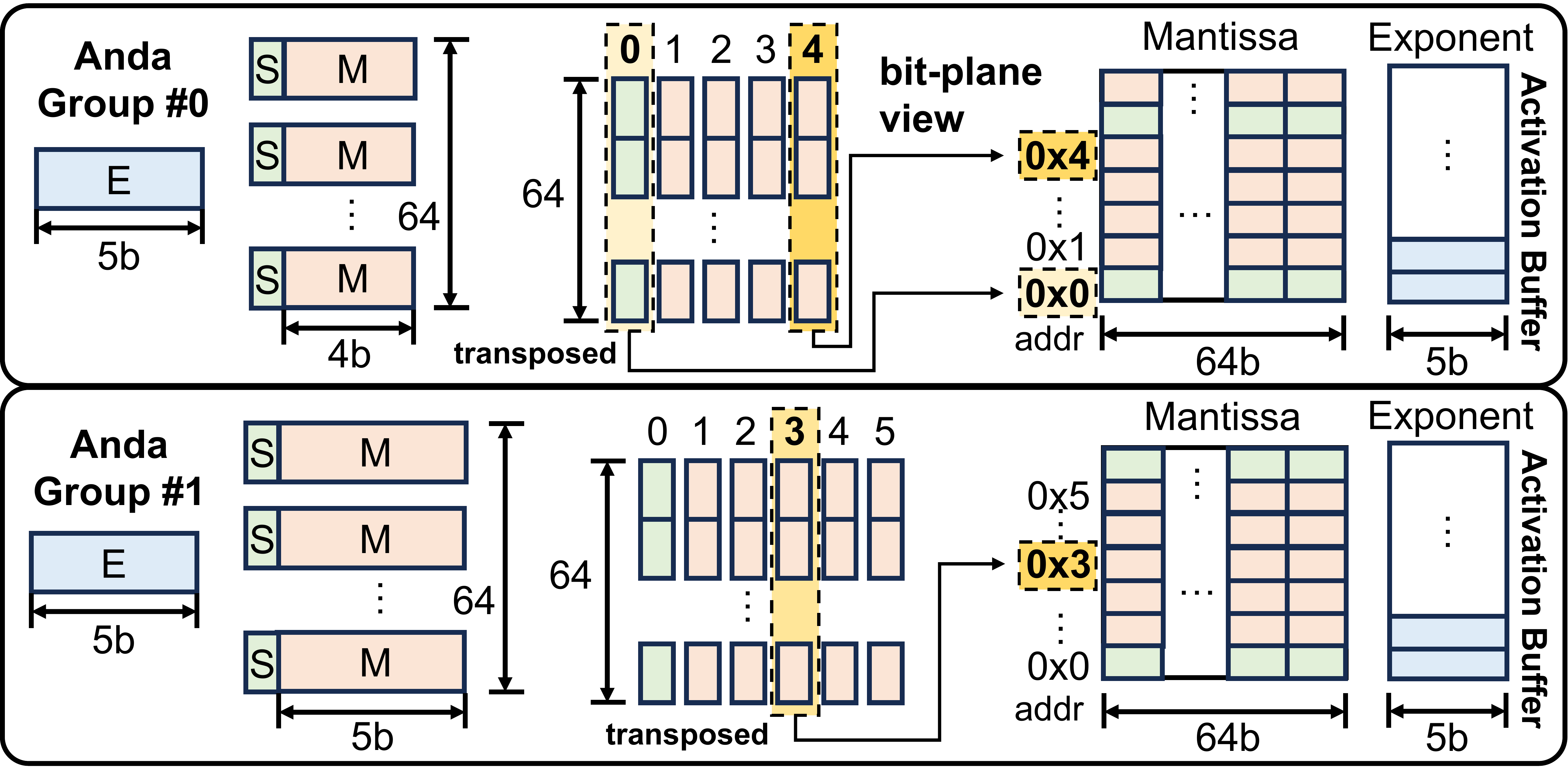}
    \caption{The proposed bit-plane data layout scheme in memory for efficient variable-length activation data storage.}
    \label{fig:mem_layout}
    \vspace{-0.5em}
\end{figure}

\subsection{Bit-plane Data Layout Scheme} \label{subsec:scheme}

\ms{Anda-based activation values feature a variable-length mantissa, necessitating careful data layout arrangement in the on-chip buffer to maintain regular memory access. Otherwise, irregular memory accesses caused by an ineffective data layout could completely undo the benefits provided by Anda.}



\rOneFC{To tackle these challenges, we propose the bit-plane data layout scheme as illustrated in Fig.~\ref{fig:mem_layout}.}
\rOneFC{\ms{Unlike prior fixed-length data arrangement methods~\cite{noh2023flexblock, kim2024isca, sharma2018bit, huang2024precision}}, \ms{which treat each FP data element as an \rg{atomic}
unit,}
our approach separates and reorganizes the sign bit, mantissa, and exponent of FP numbers within grouped data blocks from a bit-plane view.}
\rThreeFC{A transposed data arrangement~\cite{li2021geo} is introduced where bits of the same significance across multiple numbers are packed together to keep the regularity of memory access.}
\ms{Taking the common memory bank word width into account, 64 Anda-type values are grouped to implement the bit-plane data layout scheme.}
\rOneFC{As shown in Fig.~\ref{fig:mem_layout}, Group \#0 shows the layout for 4-bit mantissa Anda numbers, while Group \#1 presents the arrangement for 5-bit mantissa Anda numbers. 
\ms{\rThreeFC{The variable mantissa length only reflects on the different memory address depths, without impacting memory bandwidth utilization, and can be easily managed during address generation.}}
\rThreeFC{Hence, in both cases, the bit-plane data layout efficiently accommodates these formats with varying lengths, maintaining consistent access patterns.}}
\rThreeFC{Furthermore, the bit-plane organization inherently facilitates parallel processing, inspiring the design of a novel processing unit for the Anda data format to enhance LLM inference in both computing and energy efficiency.}

\begin{figure}[tbp]
    \centering
    \includegraphics[width=0.96\linewidth]{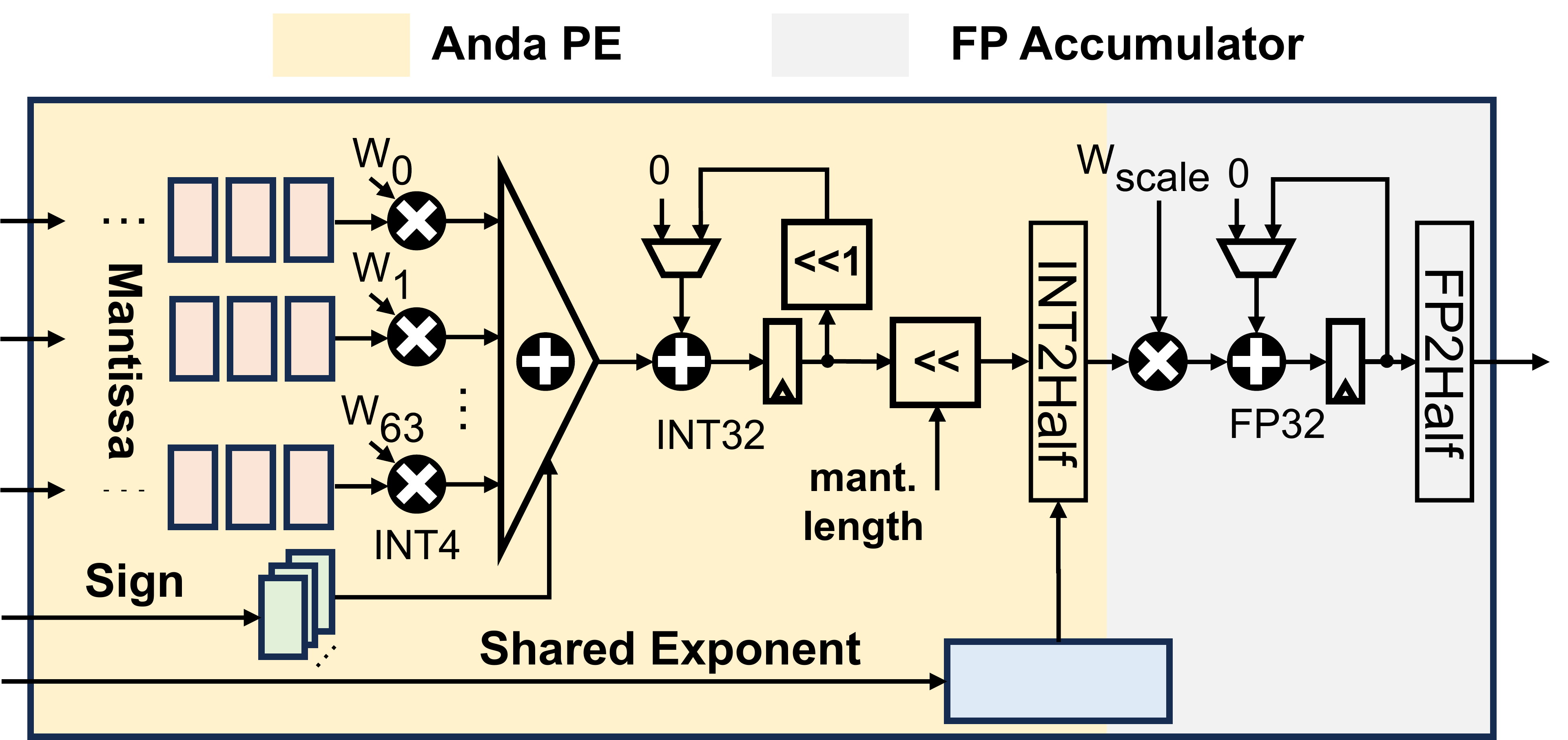}
    \caption{The architecture of Anda-enhanced bit-serial processing unit, which enables efficient dot-product operations for Anda activations and INT weights.}
    \label{fig:dpu_pattern}
    \vspace{-0.5em}
\end{figure}

\subsection{Anda-enhanced Bit-serial Processing Unit} \label{subsec:dpu}




\rOneFC{The Anda-enhanced bit-serial processing unit (APU), as depicted in Fig.~\ref{fig:dpu_pattern}, serves as the key computational element of the Anda architecture, embracing Anda processing element (PE) and an FP accumulator. Anda PE efficiently executes dot-product operations between variable-length Anda format activations and INT weights, seamlessly integrating with the bit-plane data layout scheme to enhance performance.}
\rTwoFC{The FP accumulator follows the PE to complete the APU functionality by accumulating the cross-group dot-product results.}

\rOneFC{The computation process begins with the Anda PE storing the sign and exponent in internal registers.}
\rOneFC{Concurrently, the INT weights are stored in the PE using a double-buffer design, allowing overlapped weight loading and computation to minimize loading latency.}
\rOneFC{The PE then loads the bit-plane mantissas and performs computations with the INT weights. By employing bit-serial processing of mantissas, the Anda PE can adapt to Anda format data of varying lengths without additional hardware overhead.}


\rOneFC{To further optimize hardware efficiency, the Anda PE implements a first-element-then-bit-plane reduction pattern. In this approach, a partial sum is obtained for each bit-plane by accumulating all elements within that bit-plane using an adder tree. This method reduces storage requirements by storing only one partial sum per bit-plane instead of all intermediate results. It also minimizes data movement and processing overhead by performing subsequent shift operations only on the single partial sum rather than individual elements. Furthermore, it significantly reduces hardware resource consumption by using a single shared accumulator for all bit-plane accumulations.}

\rOneFC{The bit-plane partial sums are then sequentially accumulated to complete the dot-product operation. Upon completion, the Anda PE dynamically shifts the dot-product result based on the Anda mantissa length and converts it to FP16 using the shared exponent. The result is then multiplied with the group-wise scale factor of the INT weights, followed by cross-group accumulation using the FP accumulator. Finally, the accumulated FP32 result is converted to FP16 for output.}

\subsection{On-the-fly Bit-plane Compressor} \label{subsec:comp}


\rOneFC{The bit-plane compressor (BPC) is a critical component of the Anda architecture, enabling on-the-fly conversion of FP16 activation values into the compressed Anda format. It efficiently addresses the challenges of variable-length Anda activation storage and transfer in LLM inference by processing a large number of activation values in parallel and outputting them in a bit-serial manner.}

\begin{figure}[tbp]
    \centering
    \includegraphics[width=0.98\linewidth]{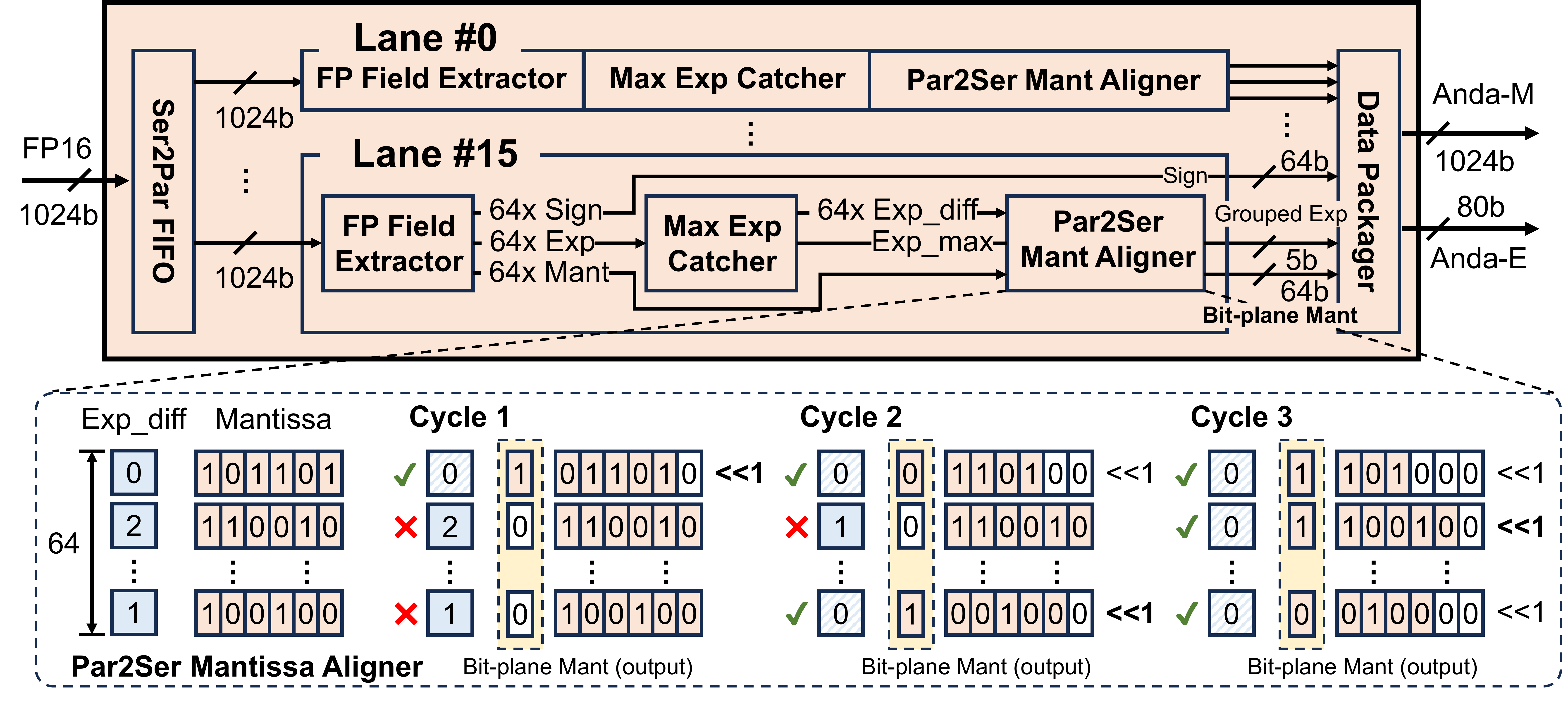}
    \caption{The architecture of the on-the-fly bit plane compressor and the mantissa alignment process performed in the parallel-to-serial mantissa aligner.}
    \label{fig:bpc}
    \vspace{-0.5em}
\end{figure}

\rOneFC{Fig.~\ref{fig:bpc} illustrates the architecture of the proposed BPC.}
\rOneFC{It consists of 16 parallel lanes, each capable of processing 64 grouped FP16 values simultaneously. Within each lane, the FP field extractor decomposes the FP16 inputs into their sign, exponent, and mantissa components. 
The maximum exponent catcher identifies the maximum exponent within a grouped lane, and then calculates the difference of each exponent to the shared maximum exponent.}

\rOneFC{The core of the compression process lies in the mantissa alignment performed by the parallel-to-serial mantissa aligner. 
\rTwoFC{As shown in Fig.~\ref{fig:bpc}, in each cycle, each element's exponent difference decreases by one until it reaches zero. When the exponent difference is zero, the most significant mantissa bit of that element should be shifted out each cycle; otherwise, it remains unchanged and output zero.}
\rTwoFC{The shifted-out bits among each element in the lane are packed into the bit-plane aligned mantissa.}
This process continues for multiple cycles until the number of output bit-planes reaches the configurable mantissa length.}
\rOneFC{The parallel-to-serial mantissa alignment process generates compressed bit-planes directly. The resulting bit-serial output, along with the sign bits and shared maximum exponents from all lanes, is passed to the data packager unit. This unit assembles the final compressed output in a format compatible with the proposed bit-plane compression scheme.}
\rThreeFC{The proposed bit-serial mantissa aligner is more area-efficient compared to existing bit-parallel aligners~\cite{zhang2022fast, jang2024figna}, requiring only a comparator and shifter. In contrast, bit-parallel designs need multiple shifters and comparators for single-cycle dynamic shifting~\cite{das2008timing}. While our bit-serial aligner introduces some latency, it can largely overlap with APU computations, with little impact on overall system performance.}

\subsection{Overall Architecture} 
\label{subsec:overall}

\begin{figure}[tbp]
    \centering
    \includegraphics[width=\linewidth]{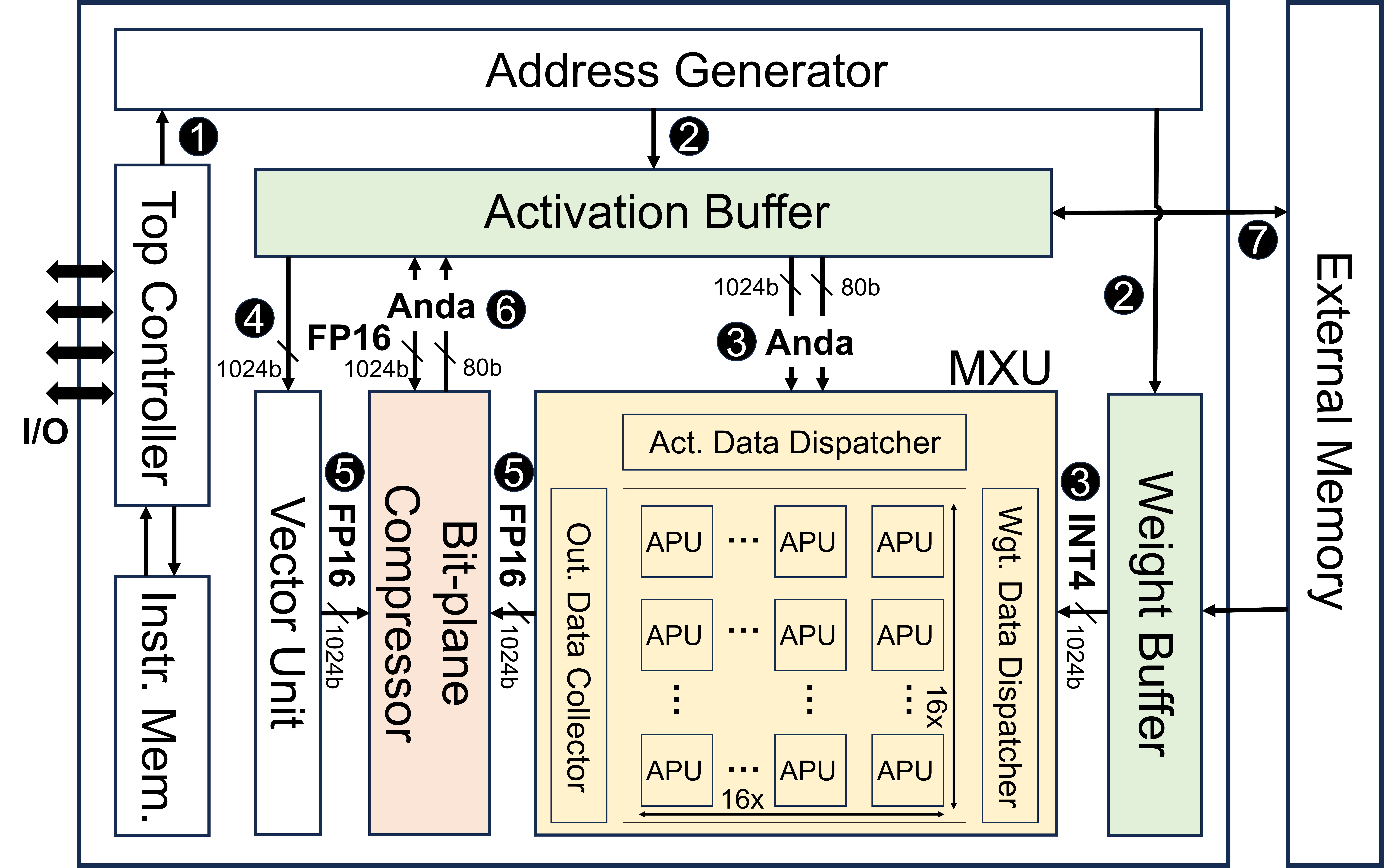}
    \caption{Anda system architecture.}
    \label{fig:arch}
\end{figure}

\rOneFC{Fig.~\ref{fig:arch} illustrates the overall architecture of Anda, which includes the top controller, address generator, activation buffer, weight buffer, matrix computation unit (MXU), vector unit, and bit-plane compressor.}
The LLM inference is orchestrated as follows: \rOneFC{\ding{182} Initially, the instruction memory is programmed through the I/O interface of the top controller, which governs the address generator during operation.}
\rOneFC{\ding{183} The address generator produces read and write addresses for both activation and weight buffers.} 
\rTwoFC{Both the activation buffer and weight buffer follow the proposed bit-plane-based data layout for efficient data handling.}
\rTwoFC{\ding{184} The MXU, featuring a 16$\times$16 APU array, performs FP-INT GeMM operations following typical output stationary dataflow~\cite{lee2024tender}.}
\rTwoFC{The weight data dispatcher, equipped with registers, allows overlapping weight loading and computation, broadcasting weights row-wise to each APU for data reuse. 
The activation data dispatcher supplies a bit-plane vector of activations each cycle, sequentially feeding it into the MXU and sharing it across columns to maximize input reuse and enable multiple calculations with the same input.}
\rTwoFC{Upon completing the GeMM computation, the output results are delivered to the BPC via the output data dispatcher.}
\rOneFC{\ding{185}~Complementing MXU, the vector unit processes the non-linear functions of the transformer block.}
\rOneFC{\ding{186}~FP16 outputs of MXU or vector unit can be optionally compressed to Anda format by the BPC, optimizing storage efficiency.}
\rOneFC{\ding{187}~Processed \ms{outputs are} 
written back to the activation buffer.
\ding{188}~Finally, activation results are transferred 
to external memory for subsequent operations.} 

\begin{table*}[htbp]
\centering
\scriptsize
\setlength{\tabcolsep}{2pt}
\begin{threeparttable}
\caption{\rThreeFC{Comparison of computation methods of weight-only quantized LLMs} \\ Key Metrics: Perplexity (Black, lower is better), \textcolor{red}{Accuracy Drop (Red)} and \textcolor[HTML]{3A9262}{BOPs Saving (Green)}}
\label{tab:algo_res}
\begin{tabular}{@{}ll|*{9}{>{\arraybackslash}p{0.09\textwidth}}@{}}
\toprule
\multicolumn{2}{c|}{} & OPT-1.3B & OPT-2.7B & OPT-6.7B & LLaMA-7B & LLaMA2-7B & OPT-13B & LLaMA-13B & LLaMA2-13B & OPT-30B \\ \midrule
\multirow{6}{*}{\rotatebox[origin=c]{90}{WikiText2}} & FP16 & ${14.62}$ & ${12.47}$ & ${10.86}$ & ${5.68}$ & ${5.47}$ & ${10.13}$ & ${5.09}$ & ${4.88}$ & ${9.56}$ \\ 
& Omniquant~\cite{shao2024omniquant} & ${14.88}^{\textcolor{red}{0.00\%}}_{\textcolor[HTML]{3A9262}{1.00\times}}$ & ${12.65}^{\textcolor{red}{0.00\%}}_{\textcolor[HTML]{3A9262}{1.00\times}}$ & ${10.96}^{\textcolor{red}{0.00\%}}_{\textcolor[HTML]{3A9262}{1.00\times}}$ & ${5.77}^{\textcolor{red}{0.00\%}}_{\textcolor[HTML]{3A9262}{1.00\times}}$ & ${5.59}^{\textcolor{red}{0.00\%}}_{\textcolor[HTML]{3A9262}{1.00\times}}$ & ${10.20}^{\textcolor{red}{0.00\%}}_{\textcolor[HTML]{3A9262}{1.00\times}}$ & ${5.17}^{\textcolor{red}{0.00\%}}_{\textcolor[HTML]{3A9262}{1.00\times}}$ & ${4.95}^{\textcolor{red}{0.00\%}}_{\textcolor[HTML]{3A9262}{1.00\times}}$ & ${9.62}^{\textcolor{red}{0.00\%}}_{\textcolor[HTML]{3A9262}{1.00\times}}$ \\ 
& FIGNA~\cite{jang2024figna} & ${14.90}^{\textcolor{red}{-0.13\%}}_{\textcolor[HTML]{3A9262}{1.23\times}}$ & ${12.65}^{\textcolor{red}{0.00\%}}_{\textcolor[HTML]{3A9262}{1.23\times}}$ & ${10.96}^{\textcolor{red}{0.00\%}}_{\textcolor[HTML]{3A9262}{1.23\times}}$ & ${5.78}^{\textcolor{red}{-0.17\%}}_{\textcolor[HTML]{3A9262}{1.23\times}}$ & ${5.60}^{\textcolor{red}{-0.18\%}}_{\textcolor[HTML]{3A9262}{1.23\times}}$ & ${10.22}^{\textcolor{red}{-0.20\%}}_{\textcolor[HTML]{3A9262}{1.23\times}}$ & ${5.18}^{\textcolor{red}{-0.19\%}}_{\textcolor[HTML]{3A9262}{1.23\times}}$ & ${4.96}^{\textcolor{red}{-0.20\%}}_{\textcolor[HTML]{3A9262}{1.23\times}}$ & ${9.61}^{\textcolor{red}{0.10\%}}_{\textcolor[HTML]{3A9262}{1.23\times}}$ \\ 
& VS-Quant\tnote{*}~\cite{keller202395} & ${19.04}^{\textcolor{red}{-27.96\%}}_{\textcolor[HTML]{3A9262}{4.00\times}}$ & ${16.41}^{\textcolor{red}{-22.91\%}}_{\textcolor[HTML]{3A9262}{4.00\times}}$ & ${12.24}^{\textcolor{red}{-11.68\%}}_{\textcolor[HTML]{3A9262}{4.00\times}}$ & ${7.45}^{\textcolor{red}{-29.12\%}}_{\textcolor[HTML]{3A9262}{4.00\times}}$ & ${8.26}^{\textcolor{red}{-47.76\%}}_{\textcolor[HTML]{3A9262}{4.00\times}}$ & ${11.63}^{\textcolor{red}{-13.98\%}}_{\textcolor[HTML]{3A9262}{4.00\times}}$ & ${6.36}^{\textcolor{red}{-23.02\%}}_{\textcolor[HTML]{3A9262}{4.00\times}}$ & ${6.43}^{\textcolor{red}{-29.90\%}}_{\textcolor[HTML]{3A9262}{4.00\times}}$ & ${10.67}^{\textcolor{red}{-10.91\%}}_{\textcolor[HTML]{3A9262}{4.00\times}}$ \\ 
& Ours (0.1\%) & \cellcolor{yellow!30}${14.91}^{\textcolor{red}{-0.20\%}}_{\textcolor[HTML]{3A9262}{2.74\times}}$ & \cellcolor{yellow!30}${12.66}^{\textcolor{red}{-0.07\%}}_{\textcolor[HTML]{3A9262}{2.91\times}}$ & \cellcolor{yellow!30}${10.95}^{\textcolor{red}{0.09\%}}_{\textcolor[HTML]{3A9262}{3.10\times}}$ & \cellcolor{yellow!30}${5.78}^{\textcolor{red}{-0.17\%}}_{\textcolor[HTML]{3A9262}{1.99\times}}$ & \cellcolor{yellow!30}${5.60}^{\textcolor{red}{-0.18\%}}_{\textcolor[HTML]{3A9262}{1.96\times}}$ & \cellcolor{yellow!30}${10.21}^{\textcolor{red}{-0.10\%}}_{\textcolor[HTML]{3A9262}{2.86\times}}$ & \cellcolor{yellow!30}${5.18}^{\textcolor{red}{-0.19\%}}_{\textcolor[HTML]{3A9262}{1.88\times}}$ & \cellcolor{yellow!30}${4.96}^{\textcolor{red}{-0.20\%}}_{\textcolor[HTML]{3A9262}{1.80\times}}$ & \cellcolor{yellow!30}${9.60}^{\textcolor{red}{0.20\%}}_{\textcolor[HTML]{3A9262}{3.10\times}}$ \\ 
& Ours (1\%) & \cellcolor{yellow!90}${14.99}^{\textcolor{red}{-0.74\%}}_{\textcolor[HTML]{3A9262}{2.95\times}}$ & \cellcolor{yellow!90}${12.76}^{\textcolor{red}{-0.86\%}}_{\textcolor[HTML]{3A9262}{3.25\times}}$ & \cellcolor{yellow!90}${10.99}^{\textcolor{red}{-0.27\%}}_{\textcolor[HTML]{3A9262}{3.31\times}}$ & \cellcolor{yellow!90}${5.82}^{\textcolor{red}{-0.87\%}}_{\textcolor[HTML]{3A9262}{2.56\times}}$ & \cellcolor{yellow!90}${5.65}^{\textcolor{red}{-1.07\%}}_{\textcolor[HTML]{3A9262}{2.56\times}}$ & \cellcolor{yellow!90}${10.30}^{\textcolor{red}{-0.98\%}}_{\textcolor[HTML]{3A9262}{3.20\times}}$ & \cellcolor{yellow!90}${5.23}^{\textcolor{red}{-1.16\%}}_{\textcolor[HTML]{3A9262}{2.59\times}}$ & \cellcolor{yellow!90}${5.00}^{\textcolor{red}{-1.01\%}}_{\textcolor[HTML]{3A9262}{2.44\times}}$ & \cellcolor{yellow!90}${9.71}^{\textcolor{red}{-0.94\%}}_{\textcolor[HTML]{3A9262}{3.31\times}}$ \\ 
\midrule 
\multirow{6}{*}{\rotatebox[origin=c]{90}{PTB}} & FP16 & ${16.96}$ & ${15.12}$ & ${13.09}$ & ${8.80}$ & ${20.82}$ & ${12.34}$ & ${8.07}$ & ${28.93}$ & ${11.84}$ \\ 
& Omniquant~\cite{shao2024omniquant} & ${17.40}^{\textcolor{red}{0.00\%}}_{\textcolor[HTML]{3A9262}{1.00\times}}$ & ${15.28}^{\textcolor{red}{0.00\%}}_{\textcolor[HTML]{3A9262}{1.00\times}}$ & ${13.25}^{\textcolor{red}{0.00\%}}_{\textcolor[HTML]{3A9262}{1.00\times}}$ & ${8.97}^{\textcolor{red}{0.00\%}}_{\textcolor[HTML]{3A9262}{1.00\times}}$ & ${21.52}^{\textcolor{red}{0.00\%}}_{\textcolor[HTML]{3A9262}{1.00\times}}$ & ${12.46}^{\textcolor{red}{0.00\%}}_{\textcolor[HTML]{3A9262}{1.00\times}}$ & ${8.14}^{\textcolor{red}{0.00\%}}_{\textcolor[HTML]{3A9262}{1.00\times}}$ & ${30.19}^{\textcolor{red}{0.00\%}}_{\textcolor[HTML]{3A9262}{1.00\times}}$ & ${11.94}^{\textcolor{red}{0.00\%}}_{\textcolor[HTML]{3A9262}{1.00\times}}$ \\ 
& FIGNA~\cite{jang2024figna} & ${17.41}^{\textcolor{red}{-0.06\%}}_{\textcolor[HTML]{3A9262}{1.23\times}}$ & ${15.28}^{\textcolor{red}{0.00\%}}_{\textcolor[HTML]{3A9262}{1.23\times}}$ & ${13.26}^{\textcolor{red}{-0.08\%}}_{\textcolor[HTML]{3A9262}{1.23\times}}$ & ${8.98}^{\textcolor{red}{-0.11\%}}_{\textcolor[HTML]{3A9262}{1.23\times}}$ & ${21.52}^{\textcolor{red}{0.00\%}}_{\textcolor[HTML]{3A9262}{1.23\times}}$ & ${12.47}^{\textcolor{red}{-0.08\%}}_{\textcolor[HTML]{3A9262}{1.23\times}}$ & ${8.15}^{\textcolor{red}{-0.12\%}}_{\textcolor[HTML]{3A9262}{1.23\times}}$ & ${30.19}^{\textcolor{red}{0.00\%}}_{\textcolor[HTML]{3A9262}{1.23\times}}$ & ${11.95}^{\textcolor{red}{-0.08\%}}_{\textcolor[HTML]{3A9262}{1.23\times}}$ \\ 
& VS-Quant\tnote{*}~\cite{keller202395} & ${25.64}^{\textcolor{red}{-47.36\%}}_{\textcolor[HTML]{3A9262}{4.00\times}}$ & ${21.09}^{\textcolor{red}{-27.55\%}}_{\textcolor[HTML]{3A9262}{4.00\times}}$ & ${15.50}^{\textcolor{red}{-16.98\%}}_{\textcolor[HTML]{3A9262}{4.00\times}}$ & ${15.78}^{\textcolor{red}{-75.92\%}}_{\textcolor[HTML]{3A9262}{4.00\times}}$ & ${53.46}^{\textcolor{red}{-148.42\%}}_{\textcolor[HTML]{3A9262}{4.00\times}}$ & ${14.47}^{\textcolor{red}{-16.13\%}}_{\textcolor[HTML]{3A9262}{4.00\times}}$ & ${12.54}^{\textcolor{red}{-54.05\%}}_{\textcolor[HTML]{3A9262}{4.00\times}}$ & ${49.91}^{\textcolor{red}{-65.32\%}}_{\textcolor[HTML]{3A9262}{4.00\times}}$ & ${13.47}^{\textcolor{red}{-12.81\%}}_{\textcolor[HTML]{3A9262}{4.00\times}}$ \\ 
& Ours (0.1\%) & \cellcolor{yellow!30}${17.41}^{\textcolor{red}{-0.06\%}}_{\textcolor[HTML]{3A9262}{1.64\times}}$ & \cellcolor{yellow!30}${15.29}^{\textcolor{red}{-0.07\%}}_{\textcolor[HTML]{3A9262}{2.13\times}}$ & \cellcolor{yellow!30}${13.26}^{\textcolor{red}{-0.08\%}}_{\textcolor[HTML]{3A9262}{2.23\times}}$ & \cellcolor{yellow!30}${8.98}^{\textcolor{red}{-0.11\%}}_{\textcolor[HTML]{3A9262}{2.21\times}}$ & \cellcolor{yellow!30}${21.48}^{\textcolor{red}{0.20\%}}_{\textcolor[HTML]{3A9262}{2.19\times}}$ & \cellcolor{yellow!30}${12.47}^{\textcolor{red}{-0.08\%}}_{\textcolor[HTML]{3A9262}{1.84\times}}$ & \cellcolor{yellow!30}${8.15}^{\textcolor{red}{-0.12\%}}_{\textcolor[HTML]{3A9262}{2.35\times}}$ & \cellcolor{yellow!30}${30.18}^{\textcolor{red}{0.03\%}}_{\textcolor[HTML]{3A9262}{2.84\times}}$ & \cellcolor{yellow!30}${11.96}^{\textcolor{red}{-0.17\%}}_{\textcolor[HTML]{3A9262}{2.23\times}}$ \\ 
& Ours (1\%) & \cellcolor{yellow!90}${17.57}^{\textcolor{red}{-0.98\%}}_{\textcolor[HTML]{3A9262}{2.31\times}}$ & \cellcolor{yellow!90}${15.42}^{\textcolor{red}{-0.91\%}}_{\textcolor[HTML]{3A9262}{2.70\times}}$ & \cellcolor{yellow!90}${13.35}^{\textcolor{red}{-0.75\%}}_{\textcolor[HTML]{3A9262}{2.86\times}}$ & \cellcolor{yellow!90}${9.05}^{\textcolor{red}{-0.89\%}}_{\textcolor[HTML]{3A9262}{2.54\times}}$ & \cellcolor{yellow!90}${21.61}^{\textcolor{red}{-0.42\%}}_{\textcolor[HTML]{3A9262}{2.39\times}}$ & \cellcolor{yellow!90}${12.58}^{\textcolor{red}{-0.96\%}}_{\textcolor[HTML]{3A9262}{2.67\times}}$ & \cellcolor{yellow!90}${8.22}^{\textcolor{red}{-0.98\%}}_{\textcolor[HTML]{3A9262}{2.67\times}}$ & \cellcolor{yellow!90}${30.41}^{\textcolor{red}{-0.73\%}}_{\textcolor[HTML]{3A9262}{2.92\times}}$ & \cellcolor{yellow!90}${12.03}^{\textcolor{red}{-0.75\%}}_{\textcolor[HTML]{3A9262}{2.86\times}}$ \\ 
\midrule 
\multirow{6}{*}{\rotatebox[origin=c]{90}{C4}} & FP16 & ${14.72}$ & ${13.16}$ & ${11.74}$ & ${7.08}$ & ${6.97}$ & ${11.20}$ & ${6.61}$ & ${6.47}$ & ${10.69}$ \\ 
& Omniquant~\cite{shao2024omniquant} & ${15.03}^{\textcolor{red}{0.00\%}}_{\textcolor[HTML]{3A9262}{1.00\times}}$ & ${13.38}^{\textcolor{red}{0.00\%}}_{\textcolor[HTML]{3A9262}{1.00\times}}$ & ${11.85}^{\textcolor{red}{0.00\%}}_{\textcolor[HTML]{3A9262}{1.00\times}}$ & ${7.21}^{\textcolor{red}{0.00\%}}_{\textcolor[HTML]{3A9262}{1.00\times}}$ & ${7.12}^{\textcolor{red}{0.00\%}}_{\textcolor[HTML]{3A9262}{1.00\times}}$ & ${11.29}^{\textcolor{red}{0.00\%}}_{\textcolor[HTML]{3A9262}{1.00\times}}$ & ${6.69}^{\textcolor{red}{0.00\%}}_{\textcolor[HTML]{3A9262}{1.00\times}}$ & ${6.56}^{\textcolor{red}{0.00\%}}_{\textcolor[HTML]{3A9262}{1.00\times}}$ & ${10.75}^{\textcolor{red}{0.00\%}}_{\textcolor[HTML]{3A9262}{1.00\times}}$ \\ 
& FIGNA~\cite{jang2024figna} & ${15.04}^{\textcolor{red}{-0.07\%}}_{\textcolor[HTML]{3A9262}{1.23\times}}$ & ${13.38}^{\textcolor{red}{0.00\%}}_{\textcolor[HTML]{3A9262}{1.23\times}}$ & ${11.86}^{\textcolor{red}{-0.08\%}}_{\textcolor[HTML]{3A9262}{1.23\times}}$ & ${7.22}^{\textcolor{red}{-0.14\%}}_{\textcolor[HTML]{3A9262}{1.23\times}}$ & ${7.12}^{\textcolor{red}{0.00\%}}_{\textcolor[HTML]{3A9262}{1.23\times}}$ & ${11.29}^{\textcolor{red}{0.00\%}}_{\textcolor[HTML]{3A9262}{1.23\times}}$ & ${6.70}^{\textcolor{red}{-0.15\%}}_{\textcolor[HTML]{3A9262}{1.23\times}}$ & ${6.57}^{\textcolor{red}{-0.15\%}}_{\textcolor[HTML]{3A9262}{1.23\times}}$ & ${10.75}^{\textcolor{red}{0.00\%}}_{\textcolor[HTML]{3A9262}{1.23\times}}$ \\ 
& VS-Quant\tnote{*}~\cite{keller202395} & ${19.00}^{\textcolor{red}{-26.41\%}}_{\textcolor[HTML]{3A9262}{4.00\times}}$ & ${16.65}^{\textcolor{red}{-19.64\%}}_{\textcolor[HTML]{3A9262}{4.00\times}}$ & ${13.13}^{\textcolor{red}{-10.80\%}}_{\textcolor[HTML]{3A9262}{4.00\times}}$ & ${8.89}^{\textcolor{red}{-23.30\%}}_{\textcolor[HTML]{3A9262}{4.00\times}}$ & ${10.36}^{\textcolor{red}{-45.51\%}}_{\textcolor[HTML]{3A9262}{4.00\times}}$ & ${12.64}^{\textcolor{red}{-11.96\%}}_{\textcolor[HTML]{3A9262}{4.00\times}}$ & ${7.85}^{\textcolor{red}{-17.34\%}}_{\textcolor[HTML]{3A9262}{4.00\times}}$ & ${8.35}^{\textcolor{red}{-27.29\%}}_{\textcolor[HTML]{3A9262}{4.00\times}}$ & ${11.91}^{\textcolor{red}{-10.79\%}}_{\textcolor[HTML]{3A9262}{4.00\times}}$ \\ 
& Ours (0.1\%) & \cellcolor{yellow!30}${15.05}^{\textcolor{red}{-0.13\%}}_{\textcolor[HTML]{3A9262}{1.86\times}}$ & \cellcolor{yellow!30}${13.40}^{\textcolor{red}{-0.15\%}}_{\textcolor[HTML]{3A9262}{2.11\times}}$ & \cellcolor{yellow!30}${11.87}^{\textcolor{red}{-0.17\%}}_{\textcolor[HTML]{3A9262}{2.23\times}}$ & \cellcolor{yellow!30}${7.22}^{\textcolor{red}{-0.14\%}}_{\textcolor[HTML]{3A9262}{2.16\times}}$ & \cellcolor{yellow!30}${7.13}^{\textcolor{red}{-0.14\%}}_{\textcolor[HTML]{3A9262}{2.04\times}}$ & \cellcolor{yellow!30}${11.30}^{\textcolor{red}{-0.09\%}}_{\textcolor[HTML]{3A9262}{2.23\times}}$ & \cellcolor{yellow!30}${6.70}^{\textcolor{red}{-0.15\%}}_{\textcolor[HTML]{3A9262}{2.16\times}}$ & \cellcolor{yellow!30}${6.57}^{\textcolor{red}{-0.15\%}}_{\textcolor[HTML]{3A9262}{2.14\times}}$ & \cellcolor{yellow!30}${10.76}^{\textcolor{red}{-0.09\%}}_{\textcolor[HTML]{3A9262}{2.31\times}}$ \\ 
& Ours (1\%) & \cellcolor{yellow!90}${15.17}^{\textcolor{red}{-0.93\%}}_{\textcolor[HTML]{3A9262}{2.43\times}}$ & \cellcolor{yellow!90}${13.51}^{\textcolor{red}{-0.96\%}}_{\textcolor[HTML]{3A9262}{2.86\times}}$ & \cellcolor{yellow!90}${11.97}^{\textcolor{red}{-1.01\%}}_{\textcolor[HTML]{3A9262}{3.05\times}}$ & \cellcolor{yellow!90}${7.28}^{\textcolor{red}{-0.97\%}}_{\textcolor[HTML]{3A9262}{2.70\times}}$ & \cellcolor{yellow!90}${7.19}^{\textcolor{red}{-0.98\%}}_{\textcolor[HTML]{3A9262}{2.67\times}}$ & \cellcolor{yellow!90}${11.37}^{\textcolor{red}{-0.71\%}}_{\textcolor[HTML]{3A9262}{2.86\times}}$ & \cellcolor{yellow!90}${6.74}^{\textcolor{red}{-0.75\%}}_{\textcolor[HTML]{3A9262}{2.70\times}}$ & \cellcolor{yellow!90}${6.62}^{\textcolor{red}{-0.91\%}}_{\textcolor[HTML]{3A9262}{2.70\times}}$ & \cellcolor{yellow!90}${10.85}^{\textcolor{red}{-0.93\%}}_{\textcolor[HTML]{3A9262}{3.09\times}}$ \\ 
\bottomrule
\end{tabular}
\begin{tablenotes}
\item[*] \rThreeFC{For fair comparison in the PTQ scenario, we directly apply VS-Quant's 4-bit data format without the costly retraining typically required by this method.}
\end{tablenotes}
\end{threeparttable}
\vspace{-1.5em}
\end{table*}

\section{Evaluation} \label{sec:eval}

\subsection{Experimental Setup}
\label{subsec:exp_setup}

\rOneFC{\textbf{LLM Benchmarks:} To demonstrate the wide applicability of our proposed method, we benchmark various open-source LLMs using PyTorch and Hugging Face libraries.}
\rOneFC{We evaluate their performance on WikiText-2~\cite{merity2017pointer}, Penn Treebank (PTB)~\cite{marcus1994penn}, and C4~\cite{raffel2020exploring} datasets. The models used in our benchmarks range in size from 1.3B to 30B parameters and are selected from OPT~\cite{zhang2022opt}, LLaMA~\cite{touvron2023llama}, and LLaMA2~\cite{touvron2023llama2} families, enabling the effectiveness assessment of our method across different model scales and architectures.}


\rOneFC{\textbf{Quantization Baselines:}
To validate the model accuracy when replacing FP activations with the proposed Anda format, we compare against the following SotA competitors:
(a) Full-precision baseline where both activations and weights are represented in FP16.
(b) Weight-only PTQ baseline using Omniquant~\cite{shao2024omniquant} with W4A16g128 scheme, which uses 4-bit weight quantization with a group size of 128. 
(c) Lossless BFP baseline that adopts FIGNA's~\cite{jang2024figna} approach of using extended mantissa lengths to maintain model accuracy.
(d) Aggressive BFP baseline that employs 4-bit mantissa to activations using VS-Quant~\cite{dai2021vs} quantization scheme. 
\rThreeFC{For fair comparison in the PTQ scenario, we directly apply VS-Quant's 4-bit data format without the typically required costly retraining.}
For baseline (c), (d), and our Anda method, we use baseline (b) as the starting point, and the group size of BFP activations is uniformly set to 64.}

\rOneFC{\textbf{Hardware Baselines:} To further highlight the advantages of the Anda scheme, we compare the Anda hardware
architecture against several SotA platforms:
(a) FP-FP: A spatial accelerator based on FP16 Tensor Cores~\cite{zadeh2022mokey}, representative of current GPU architectures~\cite{sun2022dissecting}.
(b) FP-INT: An enhanced Tensor Core-based accelerator featuring specialized FP-INT computation units.
(c) \rThreeFC{iFPU~\cite{kim2023winning}}: A spatial accelerator employs bit-serial computation for INT weights and dynamically converts FP activations to BFP format with extended mantissa before computation.
(d) FIGNA~\cite{jang2024figna}: An evolution of (c) leveraging bit-parallel computation and optimized mantissa length in BFP to enhance computational efficiency.
All systems are configured to have \rThreeFC{the same operating clock frequency of 285 MHz}, equivalent peak throughput, and on-chip memory resources to ensure a fair comparison~\cite{shi2024bitwave}.}

\rOneFC{\textbf{Evaluation Methodologies}:
Our comprehensive assessment encompasses both model accuracy and hardware efficiency\rThreeFC{, where we focus on optimizing the dominated FP-INT GeMM operations, keeping the others, e.g., KV cache~\cite{luohekeep}, in FP16}. 
For model accuracy, we employ three key metrics: (a) perplexity (PPL)~\cite{jelinek1977perplexity} using a 2048 sequence length, where lower values indicate higher model accuracy; (b) relative accuracy loss of FIGNA, VS-Quant, and our Anda method compared to Omniquant, which demonstrates the accuracy drop during BFP conversion in the deployment process; and (c) bit-level operations (BOPs)~\cite{abati2023resq} reduction, quantifying the decrease in theoretical model computation.
\rThreeFC{Here we consider one FP16-INT4 operation equivalent to 64 BOPs, 
as this approximates the bit-level computational complexity of an FP16-INT4 multiply-accumulate operation.}
Hardware performance evaluation spans two levels: at the PE level, we analyze area and power along with corresponding efficiencies; at the system level, we compare Anda with baseline architectures, focusing on speedup and energy efficiency\rThreeFC{, when performing FP-INT GeMMs where the FP activations can be drop-in replaced by Anda data format}. 
In alignment with prior studies~\cite{guo2023olive, xiao2023smoothquant, jang2024figna, kim2023winning}, system-level evaluation uses LLMs with a batch size of 1 and the maximum acceptable input sequence length under the WikiText2 dataset.
PE-level baselines and Anda system are implemented in SystemVerilog RTL and synthesized using Cadence Genus~\cite{candence2024genus} at 16nm technology node, operating at a clock frequency of 285 MHz and 0.8 V nominal voltage. 
Power evaluation is performed using value change dump (VCD) files generated from synthesized netlist simulations and analyzed by Genus. 
A cycle-accurate simulator, rigorously verified against functional simulations, assesses the energy and performance of Anda and baseline accelerators. 
HBM2 memory is modeled with an access energy of 3.9 pJ/bit and a bandwidth of 256 GB/s~\cite{jouppi2021ten}.}

\subsection{Inference Accuracy} \label{subsec:infer_acc}

\rOneFC{We evaluate inference accuracy \rThreeFC{on validation datasets} using the 
Anda format \ms{explored by} adaptive precision search algorithm across \ms{all benchmarks.} 
\rThreeFC{For each benchmark, 128 random sequences of length 2048 are sampled from the training dataset for calibration~\cite{shao2024omniquant}. We also limit the adaptive precision search algorithm to 32 iterations.}} 
\rOneFC{Note that Anda is capable of adapting mantissa bits according to the user-defined accuracy tolerance. Therefore, we report results for two accuracy constraints: 0.1\% representing minimal loss and 1\% representing acceptable loss for most scenarios.}

\begin{figure}[tb]
    \centering
    \includegraphics[width=\linewidth]{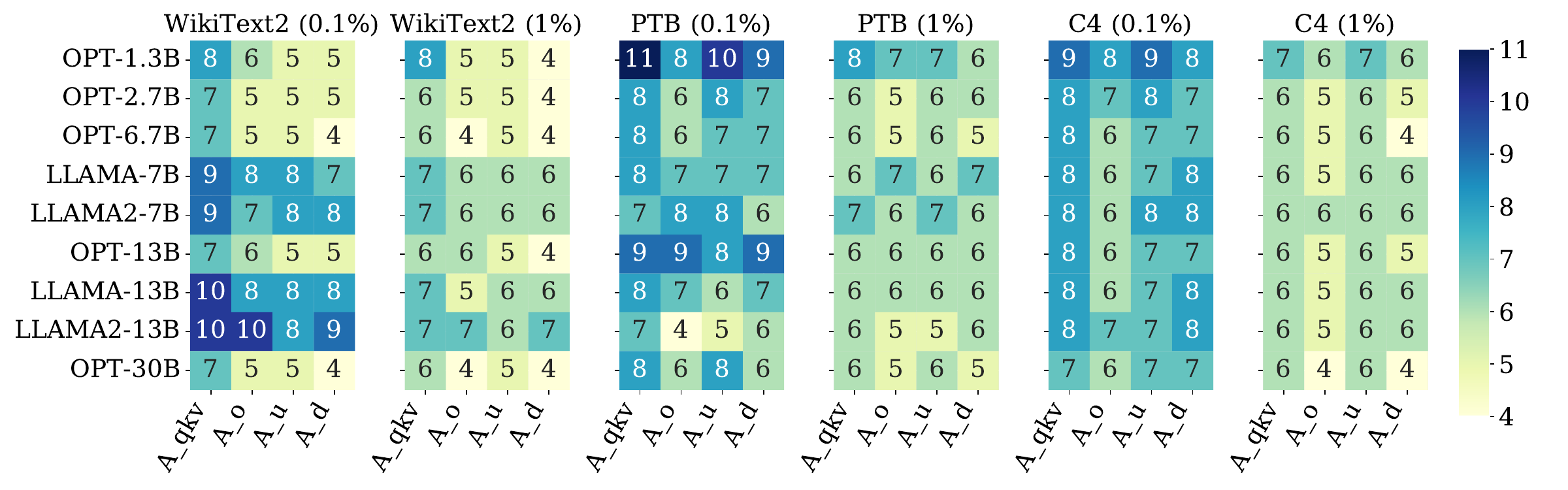}
    \caption{Identified best precision combinations of various LLMs on different datasets given different accuracy tolerances.}
    \label{fig:combination_heatmap}
    \vspace{-1ex}
\end{figure}

\rOneFC{Table~\ref{tab:algo_res} compares Anda's performance against baseline quantization methods, where PPL values are shown in black, the relative accuracy drop is displayed in red, and the BOPs reduction is presented in green.
\rTwoFC{Note that the occasional slight exceedance of the validation accuracy loss over the constraint is normal for Anda due to differences between the calibration and validation datasets.}
The results demonstrate that Anda achieves significant BOPs reductions while maintaining accuracy close to the target constraints. For instance, on the WikiText2 dataset, Anda achieves 1.80$\sim$3.10$\times$ and 2.44$\sim$3.31$\times$ BOPs reduction under 0.1\% and 1\% accuracy loss, respectively, flexibly meeting varying accuracy-efficiency requirements.}
\rOneFC{Compared to FIGNA, which yields a 1.23$\times$ BOPs reduction with shorter bit-width multiplication, Anda further achieves 1.46$\sim$2.69$\times$ BOPs reductions at similar accuracy loss levels by leveraging different mantissa lengths for activation tensors} 
\rOneFC{In contrast to VS-Quant, a BFP method requiring retraining, directly deploying it leads to severe accuracy degradation despite achieving 4$\times$ BOPs reduction. For example, VS-Quant suffers a 27.96\% accuracy loss on OPT-1.3B with WikiText2. Anda, however, achieves a much better accuracy-efficiency balance, obtaining nearly 3$\times$ BOPs reduction with only a 0.74\% accuracy loss in the same scenario.}
\rOneFC{Anda's consistent performance improvements across various models and datasets showcase the effectiveness and robust generalization of our adaptive precision search algorithm.}

\rOneFC{Fig.~\ref{fig:combination_heatmap} presents the best precision combination driven by 0.1\% and 1\% relative accuracy loss, revealing the patterns of quantization precision choices for different activation parts across models. The 
precision combinations vary under different accuracy constraints, showing the importance of the adaptive precision combination search algorithm in automatically adjusting the mantissa lengths based on LLM module sensitivity.
Examining the module types reveals that $A_{qkv}$, involved with the $Q$, $K$, and $V$ projection layers, prefers higher precision due to higher sensitivity, while $A_u$ and $A_d$ in the feed-forward layers, especially $A_d$, can be more aggressively quantized to lower precision, demonstrating higher tolerance.}


\begin{figure}[tb]
    \centering
    \includegraphics[width=\linewidth]{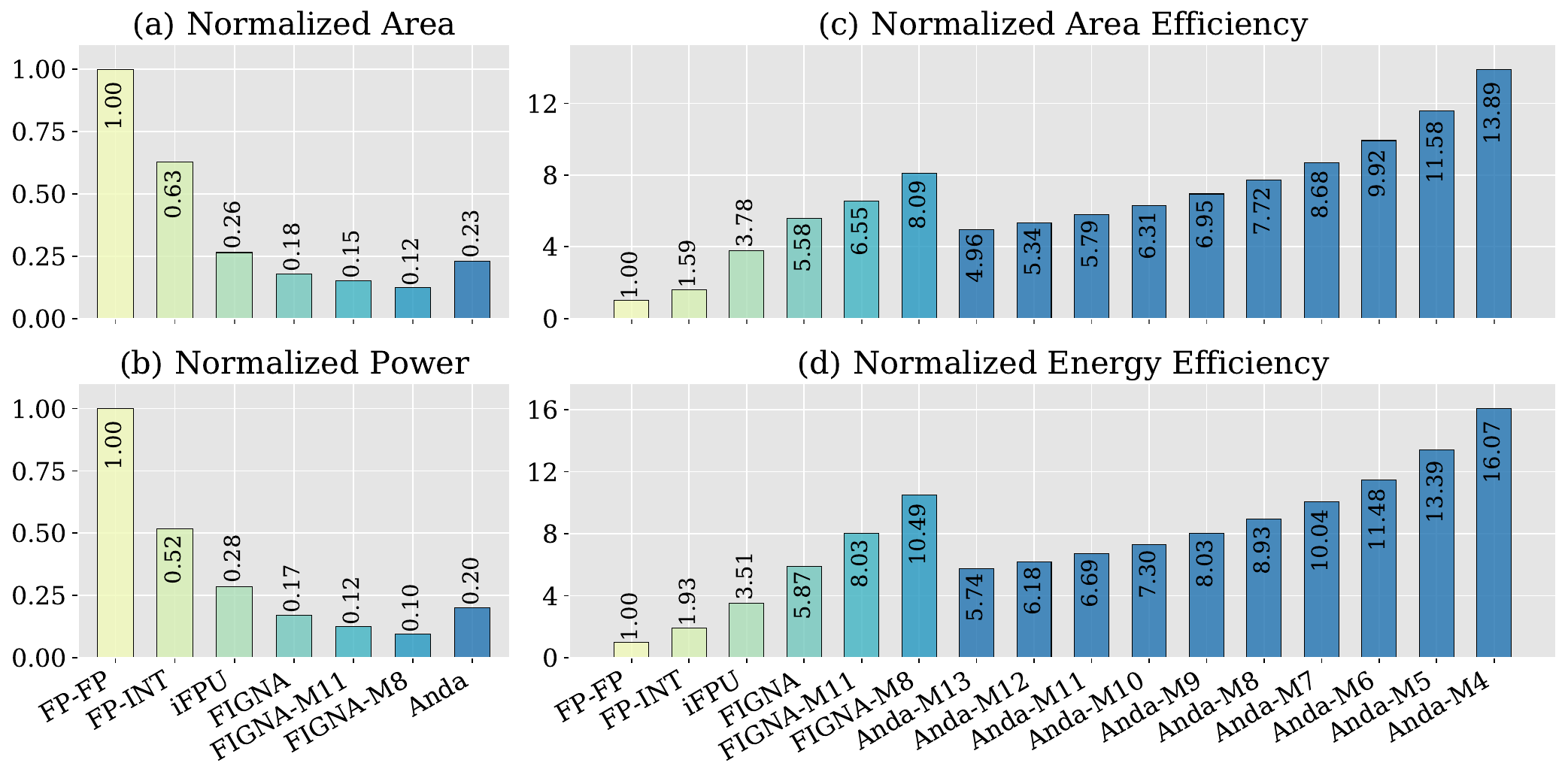}
    \caption{\rThreeFC{PE-level comparison \mvr{in terms of} area, power, area efficiency, and energy efficiency. All data are normalized to \mvr{the} GPU-like FP-FP \mvr{baseline}.}
    }
    \label{fig:pe_level}
\end{figure}
\begin{figure*}[t]
    \centering
    \includegraphics[width=\linewidth]{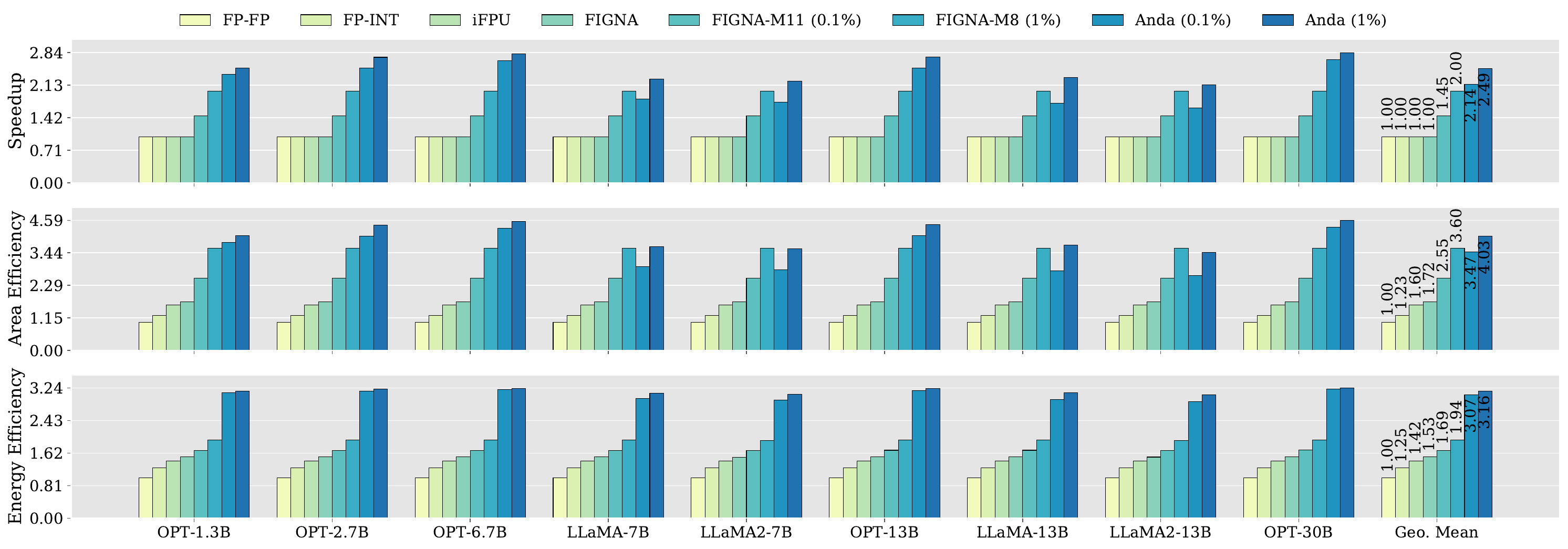}
    \caption{\rThreeFC{Speedup, area efficiency, and energy efficiency comparison across accelerators on WikiText2. All data are aligned to the GPU-like FP-FP baseline.}}
    \label{fig:benchmark}
    \vspace{-1em}
\end{figure*}




\subsection{PE-level Evaluation}

\rOneFC{We quantitatively compare the proposed Anda PE with common FP-FP units~\cite{mach2020fpnew}, enhanced FP-INT units, and dedicated PE units from iFPU~\cite{kim2023winning} and FIGNA~\cite{jang2024figna}, respectively.}
\rThreeFC{We also introduce FIGNA-M11 and FIGNA-M8 as baselines, representing bit-parallel PEs with 11-bit and 8-bit mantissas that achieve 0.1\% and 1\% accuracy degradation targets, respectively, based on the results from Fig.~\ref{fig:combination_heatmap}. Here, M(\textit{x}) denotes the number of preserved mantissa bits.}
\rOneFC{To ensure an equitable evaluation, all PEs are configured with equal computational throughput per cycle. We process an identical dot product workload across different PEs to measure area efficiency (TOPS/mm$^2$) and energy efficiency (TOPS/W).}

\rOneFC{Fig.~\ref{fig:pe_level} (a) and (b) show the area and power consumption of Anda and baseline PEs.}
\rOneFC{Anda presents significant reductions, consuming less than 60\% of the power and area compared to FP-FP and FP-INT PEs.}
\rThreeFC{This is primarily due to shared exponents, which eliminate complex alignment and normalization processes.}
\rThreeFC{Compared to iFPU~\cite{kim2023winning}, Anda offers 12\% and 29\% reductions in area and power, respectively, by avoiding high-overhead ultra-wide multipliers and registers needed for maintaining FP16 precision.}
\rThreeFC{While Anda incurs a 27\% power and 18\% area overhead compared to FIGNA due to its bit-serial structure, its adaptive precision capability can significantly reduce execution time, leading to higher efficiency.}
\rOneFC{Fig.~\ref{fig:pe_level} (c) and (d) further exhibit superior area and energy efficiency of Anda PE with variable-length mantissas.}
\rOneFC{Refering back to Fig.~\ref{fig:combination_heatmap}, the retained mantissa lengths of Anda typically range between 4$\sim$8 bits with negligible 1\% accuracy impact, resulting in the area and energy efficiency improvements of 1.38$\sim$2.48$\times$ and 1.52$\sim$2.74$\times$ over FIGNA, respectively.}
\rThreeFC{Moreover, comparing FIGNA and Anda at fixed mantissa lengths, Anda introduces some control logic overhead due to its bit-serial design.}
\rThreeFC{At 11 bits, Anda has 12\% and 17\% lower area and energy efficiency against FIGNA-M11; at 8 bits, it's 5\% and 15\% lower against FIGNA-M8. However, Anda's ability to dynamically adjust mantissa lengths based on model accuracy requirements allows it to potentially achieve higher utilization at the system level, which will be analyzed in the next subsection.}

\subsection{System-level Evaluation} \label{subsec:system}

\rThreeFC{Fig.~\ref{fig:benchmark} compares system-level speedup, area efficiency, and energy efficiency between Anda and several baselines across various LLM models.}
\rThreeFC{We also introduce bit-parallel FIGNA-M11 and FIGNA-M8 as baselines for 0.1\% and 1\% accuracy loss. Anda enables precision-scalable inference within a single hardware architecture, in contrast to FIGNA's separate implementations for each precision level.}


\rThreeFC{\textbf{Speedup:}}
\rThreeFC{Anda, utilizing the precision combinations identified in Fig.~\ref{fig:combination_heatmap}, implements scalable computation and achieves 2.14$\times$ and 2.49$\times$ speedups on average over the GPU-like FP-FP baseline at 0.1\% and 1\% accuracy loss, respectively.}
\rThreeFC{Compared to the corresponding FIGNA variants, Anda achieves 1.48$\times$ and 1.25$\times$ higher acceleration, benefiting from efficient utilization of varied mantissa precisions across tensor types.}

\rThreeFC{\textbf{Area Efficiency:}}
\rThreeFC{Anda improves area efficiency by 3.47$\times$ and 4.03$\times$ over the GPU-like FP-FP baseline at 0.1\% and 1\% loss, respectively, due to two factors: (a) shared exponent design simplifies alignment operations, improving computational unit efficiency; (b) bit-serial design fully utilizes mantissa widths of different tensor types.}
\rThreeFC{Notably, in 1\% loss with LLaMA models, FIGNA-M8's area efficiency rivals or slightly exceeds Anda due to its alignment with 8-bit precision, where bit-parallel designs excel. However, Anda’s scalable computation outperforms FIGNA by adopting more aggressive bit-widths in OPT models.}

\begin{figure}[tbp]
    \centering
    \includegraphics[width=0.96\linewidth]{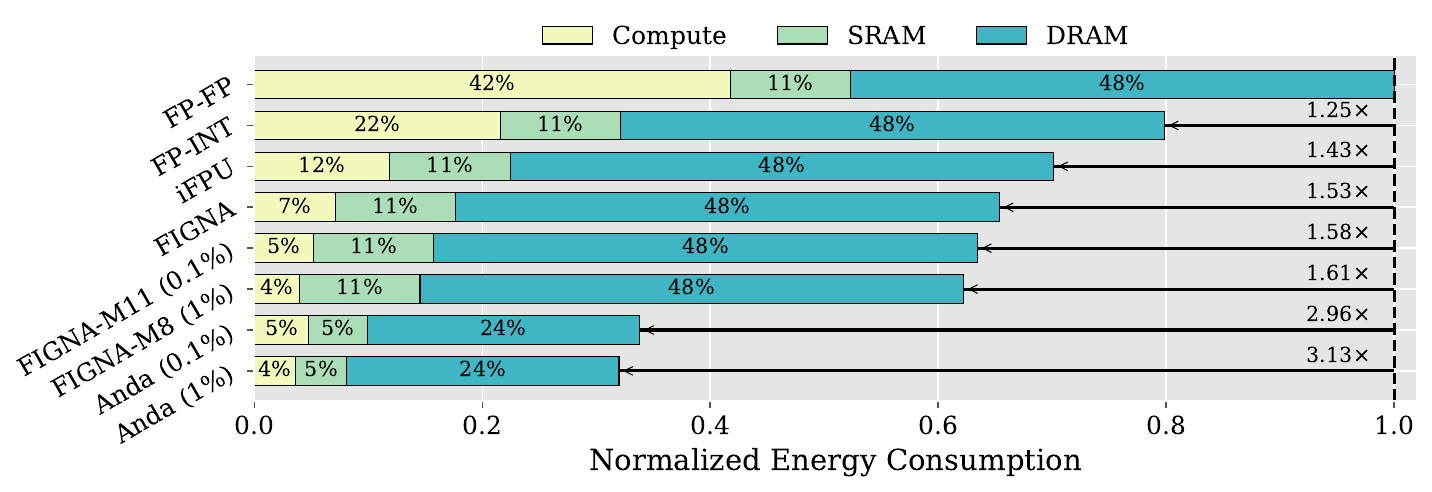}
    \caption{\rThreeFC{Energy breakdown of Anda in contrast with the baseline accelerators. Energy consumption during the LLaMA-13B inference is evaluated.}}
    \label{fig:energy_breakdown}
\end{figure}

\rThreeFC{\textbf{Energy Efficiency:}}
\rThreeFC{Anda achieves a 3.07$\times$ improvement over the GPU-like FP-FP baseline at 0.1\% accuracy loss, increasing to 3.16$\times$ at 1\% loss tolerance.}
\rThreeFC{Unlike iFPU~\cite{kim2023winning} and FIGNA~\cite{jang2024figna}, which solely optimize energy during computation, Anda’s bit-serial architecture skips redundant mantissa bit calculations to improve computational utilization, and the BPC compresses output, reducing memory access. FIGNA-M11 and FIGNA-M8 use reduced mantissa bit-parallel designs to improve computational efficiency, but rely on FP16 storage, leading to frequent data conversions, which offsets energy gains.}
\rThreeFC{Fig.~\ref{fig:energy_breakdown} further presents that compared to the GPU-like FP-FP baseline on the LLaMA-13B model, Anda reduces energy consumption by 90\%, 54\%, and 50\% for computation, SRAM, and DRAM access, respectively. 
While FIGNA achieves similar compute efficiency, Anda's architecture avoids redundant computations and FP-to-BFP conversion, reducing energy further. Moreover, Anda’s bit-plane storage scheme and BPC compression reduce memory access overhead, improving SRAM and DRAM energy efficiency by 2.2$\times$ and 2.0$\times$ compared to FIGNA.} 

\begin{table}[tb]
\centering
\caption{Area and Power Characteristics of Anda}
\label{tab:area_breakdown}
\resizebox{0.48\textwidth}{!}{%
\begin{tabular}{@{}llcc@{}}
\toprule
Component               & Setup                         & Area {[}mm$^2${]}        & Power {[}mW{]}            \\ \midrule
MXU & 16$\times$16 APUs & 0.41 (18.89\%)           & 54.34 (66.94\%)           \\
BPC    & 16 Lanes                      & 0.07 (3.23\%)            & 1.06 (1.31\%)             \\
Vector Unit             & 64 FPUs                       & 0.05 (2.30\%)            & 0.87 (1.07\%)             \\
Activation Buffer       & 1MB (Mant.) + 0.125MB (Exp.)  & 0.87 (40.09\%)           & 16.94 (20.87\%)           \\
Weight Buffer           & 1MB                           & 0.80 (36.87\%)           & 7.96 (9.81\%)             \\
Others                  & Top controller                & 0.01 (0.46\%)            & 0.01 (0.00\%)             \\ \midrule
\textbf{Total}          & \textbf{}                     & \textbf{2.17 (100.00\%)} & \textbf{81.18 (100.00\%)} \\ \bottomrule
\end{tabular}%
}
\end{table}

\begin{figure}[tp]
    \centering
    \includegraphics[width=\linewidth]{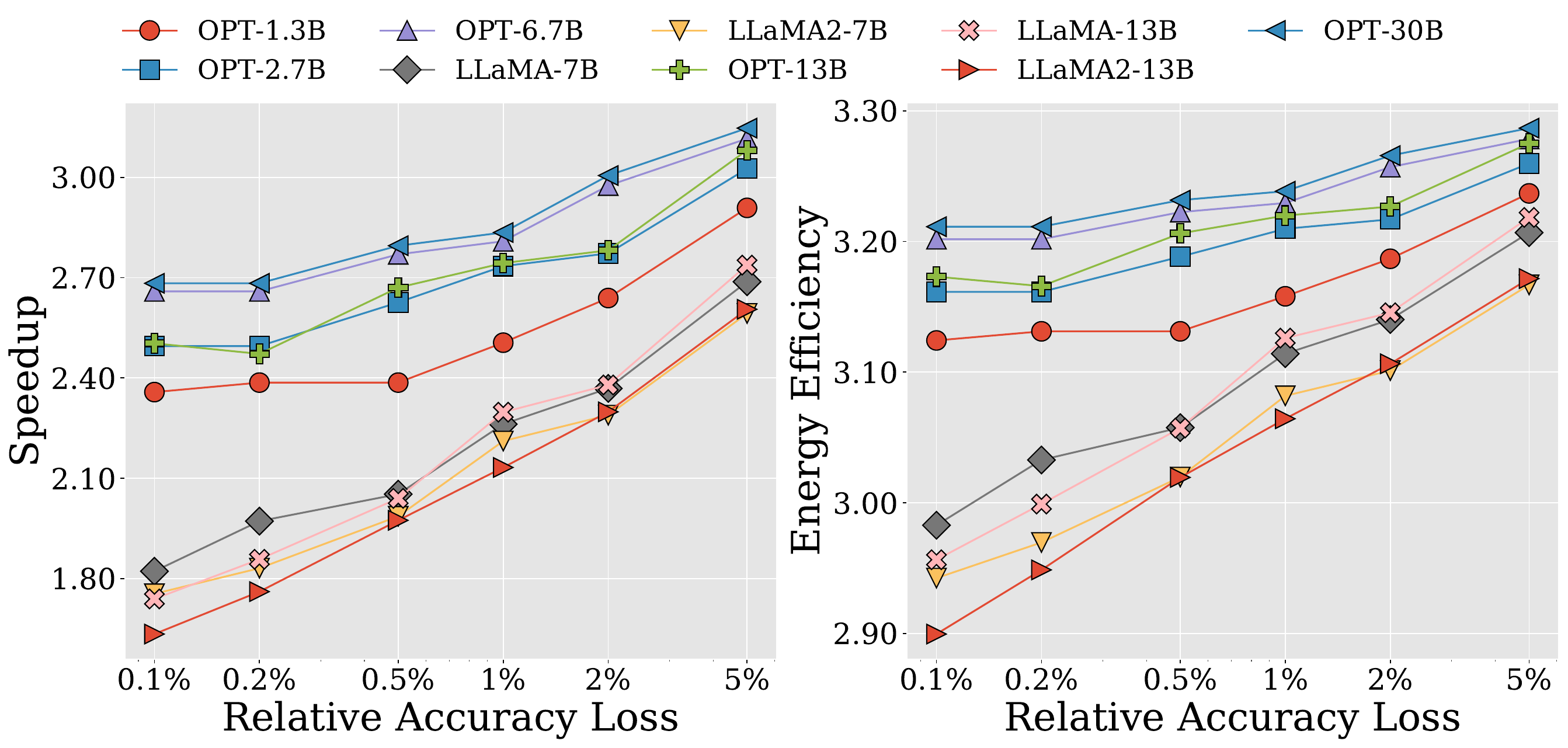}
    \caption{Speedup and energy efficiency improvement of Anda over FP-FP baseline towards various acceptable accuracy losses.}
    \label{fig:acc_eff_trade_off}
\end{figure}
\subsection{Power and Area Breakdown}
\rOneFC{We conduct a detailed hardware analysis of the Anda architecture for LLaMA-13B inference within 1\% accuracy loss. Table~\ref{tab:area_breakdown} presents the area breakdown and power distribution.}
\rOneFC{Operating at 285 MHz and 0.8 V, Anda occupies 2.17 mm$^2$ with 81.18 mW power consumption.}
The MXU, serving as the core computing component of the Anda architecture, consumes 66.94\% of the total power despite occupying only 18.89\% of the area. 
The BPC unit, which enables efficient online compression from the full-precision FP outputs to the Anda format, costs a small portion of the total area (3.23\%) and power consumption (1.31\%).
On-chip SRAM is the primary area contributor, with the activation buffer and weight buffer accounting for 40.09\% and 36.87\% of the total area, respectively. Their power consumption ratios are relatively low at 20.87\% and 9.81\% because of efficient data reuse within the Anda system.

\subsection{Accuracy-Performance Trade-off}
\rOneFC{This section explores speedup and energy efficiency improvements of the Anda system over the FP-FP baseline with accuracy loss constraints ranging from 0.1\% to 5\%.}
\rOneFC{As shown in Fig.~\ref{fig:acc_eff_trade_off}, using LLaMA-13B as an example, Anda achieves a 1.73$\times$ speedup and 2.95$\times$ energy efficiency improvement with only 0.1\% accuracy loss, increasing to 2.74$\times$ and 3.22$\times$, respectively, when the constraint is relaxed to 5\%. All models exhibit significant acceleration and efficiency gains as the tolerated accuracy loss increases.}
\rOneFC{Notably, OPT and LLaMA models exhibit distinct characteristics when using the Anda format.}
\rOneFC{This stems from OPT's lower sensitivity to bit-width reductions, allowing the use of shorter mantissa bit-widths with minimal accuracy sacrifice. Consequently, under tighter accuracy constraints, e.g., 0.1\%$\sim$0.5\%, OPT models achieve greater speedups and energy efficiency improvements compared to LLaMA models. However, as accuracy constraints relax, their performance gains gradually converge.}
\rOneFC{By integrating the adaptive precision combination search algorithm with the Anda format, our architecture achieves flexible balancing of system performance and accuracy across diverse practical application scenarios, enabling efficient LLM inference under different LLM architectures and varying requirements.}
\section{Related Works and Discussions} \label{sec:rel}






\rThreeFC{\textbf{Bit-serial \& bit-parallel computing.}}
\rThreeFC{Bit-serial computing \cite{judd2016stripes, albericio2017bit, lu2021distilling, li2022bitcluster, chang2023general, shi2024bitwave, wang2024bsvit} has long been explored in DNN acceleration, offering flexibility for variable precision computations. However, most prior work focuses on INT operations, limiting applicability to LLMs with FP activations.}
\rThreeFC{Approaches like Bitlet~\cite{lu2021distilling} and Bitlet-X~\cite{chang2023general} explore FP-based bit-serial computing but introduce complex hardware and dataflow designs due to bit-interleave schemes. In contrast, our Anda simplifies alignment overhead using variable-length grouped activation encoding, leading to a more efficient hardware design.}
\rThreeFC{Although bit-serial computing typically has lower area efficiency and higher latency than bit-parallel approaches~\cite{sharma2018bit, zhang2022fast, kim2024isca, noh2023flexblock} due to complex timing control logic, it offers higher utilization across precision-scalable scenarios.}
\rThreeFC{While Anda uses bit-serial units, its design principles can benefit bit-parallel computing as well. For instance, the proposed bit-precision combination search method can rapidly determine the required precision for bit-parallel applications, potentially improving efficiency while maintaining accuracy.}




\rThreeFC{\textbf{PTQ \& quantization-aware training (QAT).}}
\rThreeFC{Quantization, a key compression technique, is generally categorized into PTQ and QAT.
In the LLM era, PTQ~\cite{shao2024omniquant, xiao2023smoothquant, lin2024awq, frantar2023optq} is more popular, efficiently producing deployable models with good accuracy in hours on a single GPU.
In contrast, 
QAT for LLMs~\cite{liu2023llm, tseng2024quip, egiazarian2024extreme, chen2024efficientqat}, while potentially more accurate, is often impractical due to its extensive computational demands, often requiring multi-GPU systems and hundreds of training hours.
Anda adopts the PTQ approach, swiftly allocating mantissa lengths for FP activations, and can be integrated into existing deployment pipelines with minimal overhead. Future research could explore using Anda for QAT, potentially enhancing accuracy while reducing computational costs.}



\rThreeFC{\textbf{KV cache optimization.}}
\rThreeFC{In long-context scenarios, KV cache~\cite{yuan2024kv}, which stores attention keys and values during generation to avoid re-computation, becomes a memory and speed bottleneck as its size grows linearly with the number of tokens.}
\rThreeFC{Various techniques has been proposed to tackle this challenge including quantization~\cite{liu2024kivi, hooper2024kvquant}, eviction~\cite{zhang2024h2o, chen2024nacl}, sliding window~\cite{xiao2024efficient, duanmu2024skvq}, and merging strategies~\cite{zhang2024cam, kim2024compressed}. 
Anda, while focusing on FP activation compression, could synergize with these KV cache optimizations to significantly accelerate long-context LLM inference.}

\section{Conclusion} \label{sec:concls}


\rOneFC{This work presents Anda, a variable-length grouped activation data format that addresses energy and performance bottlenecks of weight-only quantized large language model (LLM) inference by exploiting redundancy in floating point activations across different models and their modules.}
To fully harness the potential of Anda, we develop an iterative post-training algorithm that optimizes bit-width allocation across LLM modules, balancing accuracy, energy efficiency, and inference speed.
We design complementary hardware optimizations to maximize the benefits of Anda, including a bit-plane-based data organization scheme in memory, Anda-enhanced bit-serial processing units, and a runtime bit-plane compressor. 
Our evaluations show that Anda achieves a 2.4$\times$ speedup\rThreeFC{, a 4.0$\times$ enhancement in area efficiency,} and a 3.1$\times$ improvement in energy efficiency on average for popular LLMs compared to 
\rThreeFC{the GPU-like FP-FP baseline.}
With its adaptability across various application scenarios and performance requirements, Anda enables efficient LLM inference in diverse deployment environments, paving the way for broader adoption of LLMs in resource-constrained settings.

\section*{Acknowledgements}
This project has been partly funded by the National Key R\&D Program of China under Grant 2022YFB4400600, the European Research Council (ERC) under grant agreement No. 101088865, the Flanders AI Research Program, the KU Leuven Hercules program, and the China Scholarship Council Program (Grant No. 202306190235). We would like to sincerely thank Mengzhao Chen at HKU, Zhe Wang at HONOR, Zhi Zhang at SenseTime for the helpful discussion, and the anonymous reviewers for their constructive feedback.


\bibliographystyle{IEEEtranS}
\bibliography{ref}








\end{document}